\documentclass[prx,epsfigure,twocolumn,superscriptaddress]{revtex4-2}
\usepackage[colorlinks=true,linkcolor=blue,urlcolor=blue,citecolor=blue,pdfusetitle]{hyperref}
\usepackage[utf8]{inputenc}
\usepackage[english]{babel}
\usepackage{amsmath}
\usepackage[caption = false]{subfig}
\usepackage{graphicx,epstopdf}
\usepackage{blindtext}
\usepackage{lipsum}
\usepackage{amsfonts}
\usepackage{bbm}
\usepackage{amssymb}
\usepackage{enumerate}
\usepackage{color}
\usepackage{latexsym}
\usepackage{physics}
\usepackage{times,txfonts}
\usepackage{siunitx}
\usepackage{subdepth}
\usepackage{upgreek}
\usepackage{tabularx}
\usepackage{tikz}

\newcommand\setrow[1]{\gdef\rowmac{#1}#1\ignorespaces}
\newcommand\clearrow{\global\let\rowmac\relax}
\clearrow

\newcommand{\myrightxarrow}[1]{\mathrel{\raisebox{-2.5pt}{$\xrightarrow{#1}$}}}

\allowdisplaybreaks

\usepackage{tikz}
\usetikzlibrary{shapes.geometric, arrows}
\tikzstyle{process} = [rectangle, minimum width=3cm, minimum height=1cm, text centered, draw=white,fill=white]
\tikzstyle{arrow} = [thick,->,>=stealth]

\makeatletter
    \renewcommand\@make@capt@title[2]{%
     \@ifx@empty\float@link{\@firstofone}{\expandafter\href\expandafter{\float@link}}%
      {\textrm{#1}}\@caption@fignum@sep#2\quad}%
    \makeatother
   
\makeatletter 
\renewcommand{\fnum@figure}{\textrm{FIG.~\thefigure}}
\makeatother

\newcommand{\UFSCar}{Departamento de Física, Universidade Federal de São Carlos, \\Rodovia Washington Luís, km 235 - SP-310, 13565-905 São Carlos, SP, Brazil}
\newcommand{\SU}{Department of Physics, Stockholm University, SE-106 91 Stockholm,
Sweden}
\newcommand{\Nice}{Universit\'e C\^ote d'Azur, CNRS, Institut de Physique de Nice, 06560 Valbonne, France}
\newcommand{\CSIC}{Instituto de Física Fundamental (IFF), Consejo Superior de Investigaciones Científicas (CSIC), \\ Calle Serrano 113b, 28006 Madrid, Spain}

\usepackage{orcidlink}

\definecolor{MyRed}{RGB}{255, 102, 102}
\definecolor{Myblue}{RGB}{102, 140, 255}

\begin{document}
\title{Phononic bright and dark states: Investigating \\ multi-mode light-matter interactions with a single trapped ion}

\author{Harry Parke~\orcidlink{0000-0001-6120-5470}}
\email{harry.parke@fysik.su.se}
\thanks{\quad These two authors contributed equally.} 
\affiliation{\SU}

\author{Robin~Thomm~\orcidlink{0009-0000-8105-8690}}
\email{robin.thomm@fysik.su.se} 
\thanks{\quad These two authors contributed equally.} 
\affiliation{\SU}

\author{Alan C. Santos~\orcidlink{0000-0002-6989-7958}}
\email{ac\_santos@iff.csic.es}
\affiliation{\SU}
\affiliation{\UFSCar}
\affiliation{\CSIC}

\author{André Cidrim~\orcidlink{0000-0003-0007-2330}}
\affiliation{\SU}
\affiliation{\UFSCar}

\author{Gerard~Higgins~\orcidlink{0000-0003-0946-8067}}
\affiliation{\SU}
\affiliation{Department of Microtechnology and Nanoscience (MC2), Chalmers University of Technology, SE-4DS 96 Gothenburg, Sweden}
\affiliation{Institute for Quantum Optics and Quantum Information (IQOQI), Austrian Academy of Sciences, A-1090 Vienna, Austria}

\author{Marion Mallweger~\orcidlink{0009-0009-8142-2988}}
\affiliation{\SU}

\author{Natalia Kuk~\orcidlink{0009-0003-0601-0314}}
\affiliation{\SU}

\author{Shalina Salim}
\affiliation{\SU}

\author{Romain Bachelard~\orcidlink{0000-0002-6026-509X}} 
\affiliation{\UFSCar}
\affiliation{\Nice}

\author{Celso J. Villas-Boas~\orcidlink{0000-0001-5622-786X}} 
\email{celsovb@df.ufscar.br}
\affiliation{\UFSCar}

\author{Markus Hennrich~\orcidlink{0000-0003-2955-7980}} 
\email{markus.hennrich@fysik.su.se}
\affiliation{\SU}

\begin{abstract}

\noindent Interference underpins some of the most practical and impactful properties of both the classical and quantum worlds. In this work we experimentally investigate a new formalism to describe interference effects, based on collective states which have enhanced or suppressed coupling to a two-level system. We employ a single trapped ion, whose electronic state is coupled to two of the ion's motional modes in order to simulate a multi-mode light-matter interaction. We observe the emergence of phononic bright and dark states for both a single phonon and a superposition of coherent states and demonstrate that a view of interference which is based solely on their decomposition in the collective basis is able to intuitively describe their coupling to a single atom. This work also marks the first time that multi-mode bright and dark states have been formed with the bounded motion of a single trapped ion and we highlight the potential of the methods discussed here for use in quantum information processing.  
\end{abstract}

\maketitle


\section{Introduction}

Atom-light interactions play a pivotal role in various forms of quantum technology. For quantum information processing, interactions with single field modes are routinely employed in systems of trapped ions \cite{Leibfried:03}, and neutral atom arrays \cite{Henriet:20} in order to directly manipulate qubits, mediate multi-qubit interactions \cite{Cirac:01, jaksch:00} or transfer information from one physical system to another \cite{Duan:01, Blinov:01, Tey:01}. Multi-mode operations can also be used for gate schemes \cite{Sorensen:02} and to apply coherent displacement and squeezing operations in trapped ion systems \cite{Meekhof:01}. In cavity QED experiments, the strong coupling regime has been shown to mitigate decoherence effects \cite{Putz:01}, enhance on-demand single photon production \cite{Walker:01, Kuhn:01, Zhang:07} and provide effective cooling of atomic ensembles \cite{Schleier-smith:01}. As the architectures for these systems continue to improve, research beyond single-mode interactions is likely to become increasingly prevalent.

A multi-mode atom-light interaction can be simulated via the coupling between a trapped ion's internal electronic states and its external quantized motion. Since phononic operations offer a high degree of tunability and control on the level of single quanta, they have already allowed such systems to simulate a range of quantum and semiclassical behaviours. For a single ion interacting with a single motional mode the Jaynes-Cummings model \cite{Jaynes:01} can be implemented to study atom-cavity dynamics \cite{Blockley:01, Cirac:02}, spin-boson models \cite{Porras:01}, and strongly correlated systems \cite{Porras:02, Deng:01}. With a chain of ions the movement of local phonons can be used to study the Bose-Hubbard model \cite{Haze:01, Debnath:01}, quantum walks \cite{Tamura:01} and quantum heat transport \cite{Bermudez:01}. Multi-ion systems have also been shown to exhibit strong phonon-mediated interactions and undergo the same superradiant phase transitions \cite{Safavi-Naini:01, Jaako:01} as have been observed in other physical systems of superconducting qubits \cite{Mlynek:01} and atomic ensembles \cite{Goban:01, Zhiqiang:01, Nagy:01}. 

When studying atom-boson interactions for two or more field modes, the total coupled system is governed by interference at the quantum level. The way in which this interference can be treated then differs depending on the occupation of each mode. For coherent states the quantized field is treated analogously to a classical wave; interference between two fields is simply a summation, and coupling to matter is determined by the average field at the object's position. However, as we move away from the boundary between classical and quantum physics, we also move beyond this intuitive treatment. To describe the interference of highly non-classical states, such as Fock states, their interaction with matter is instead attributed to quantum fluctuations of the field. 

In this work we present a novel interpretation of interference, in which both Fock states and coherent states can be treated in the same manner. In doing so we will introduce a Dicke-like bosonic basis, as described in \cite{Maximo:01}, in which bright and dark states naturally emerge. Here the term ``bright'' refers to a state in which the atom-boson coupling is enhanced compared to a single-mode state with the same occupation number and ``dark'' refers to a state in which the coupling is completely suppressed. These states are therefore defined by the interaction between multiple bosonic modes and a single two-level system.

Using a single trapped $^{88}$Sr$^{+}$ ion, we first investigate the influence of bright and dark states in the quantum regime by preparing collective states using a single phonon. We then expand our study to the classical boundary by preparing superpositions of coherent states. Theoretical predictions are confirmed by observing how bright and dark phononic states interact with a single electronic transition and we demonstrate that a non-zero atom-boson coupling is best characterized by the contribution of bright states in the collective basis. Finally, we explore how these results relate to the concept of quantum interference, providing an intuitive understanding of the effects observed.

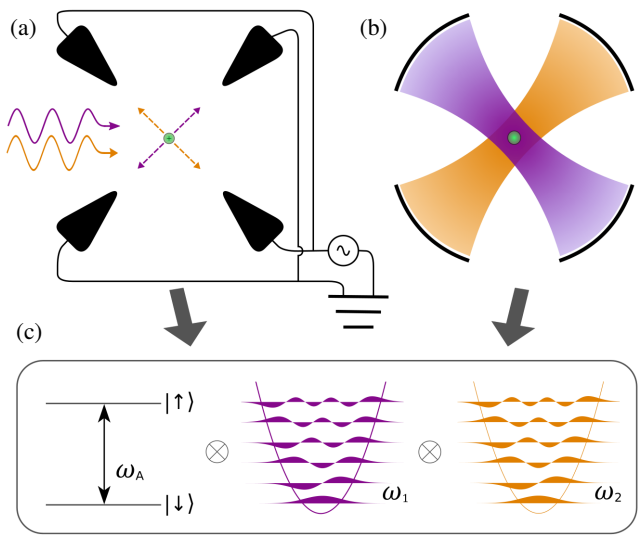
\begin{figure}[t!]
    \centering
        \includegraphics[width=\columnwidth]
        {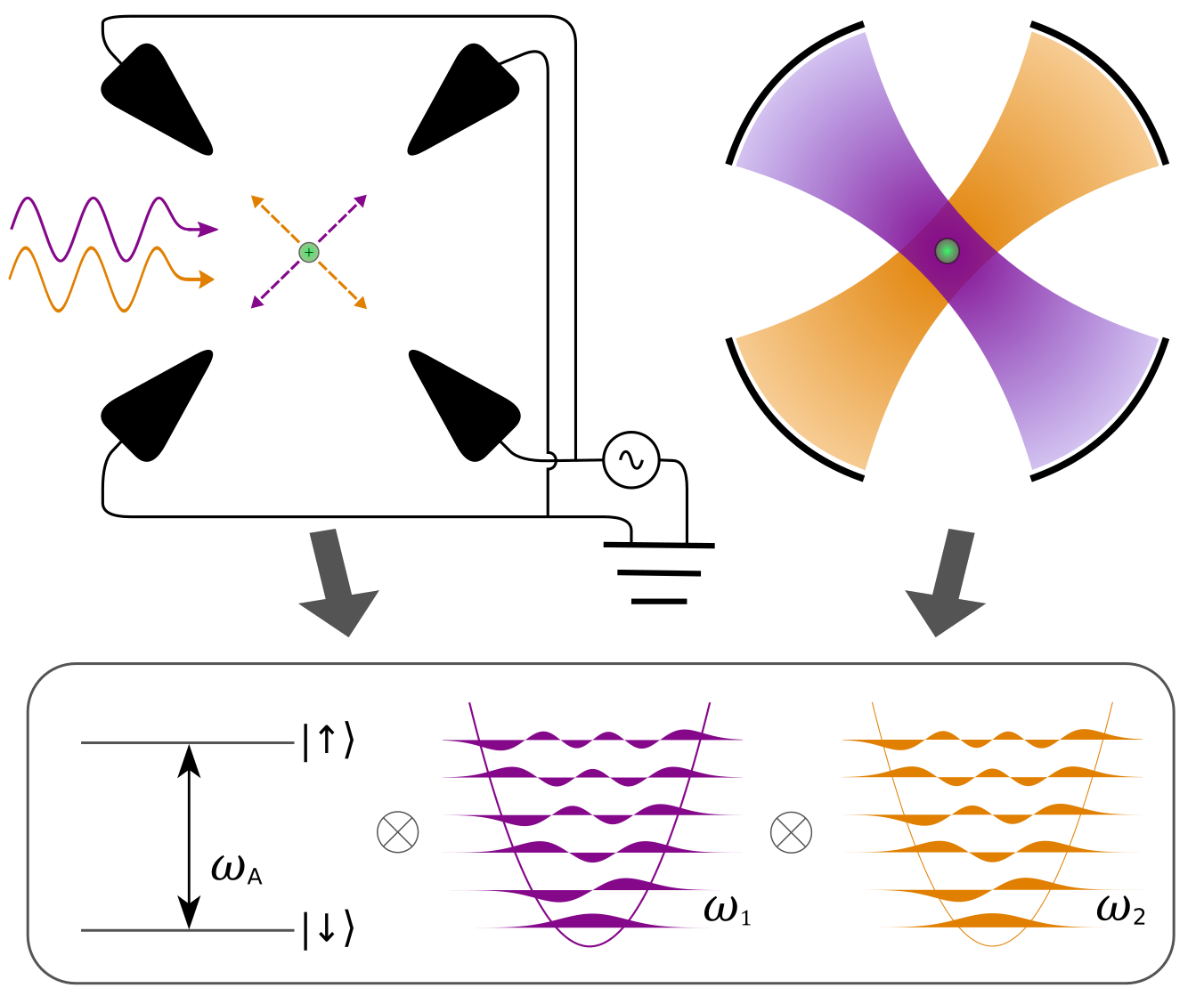}
        \begin{picture}(0,0) 
        \begin{picture}(0,0) 
            \put(-120,200){(a)}       
        \end{picture}
        \begin{picture}(0,0) 
            \put(10,200){(b)}       
        \end{picture}
            \put(-120,85){(c)}       
        \end{picture}
        \caption{Systems featuring multiple bosonic modes coupled to a single two-level system can be used to study the emergence of bright and dark states. 
        (a) Schematic of the linear Paul trap used in this work. A bichromatic laser beam simultaneously couples the two radial motional modes of a single trapped ion to the electronic degree of freedom. 
        (b) An atom in a crossed-cavity setup follows the same dynamics when interacting simultaneously with two light modes. 
        (c) Representation of the dominant terms of the interaction hamiltonian; two harmonic oscillators which are isolated from one another are independently coupled to a single two-level system.}
        \label{fig:Schematics}
\end{figure}


\section{Bright and dark phononic states in trapped ions}
The confinement of an ion in the harmonic potentials of a linear Paul trap provides an accessible simulation of atom-cavity dynamics where the interaction is not limited to a single bosonic mode, as shown in Fig.~\ref{fig:Schematics}. The 3D motion of an ion in such a trap is characterized by three individual harmonic oscillators, one along the trap axis and two in the radial plane perpendicular to it. The collective states are realized by the two radial motional modes. We apply a DC quadrupole field to offset the trapping field in one direction, spectrally separating the two modes by  $\Delta = \SI{70}{\kilo\hertz}$ so that we can individually address them. The frequencies of the two modes are $\omega_{1} = 2\pi \times \SI{1.43}{\mega\hertz}$ and $\omega_{2} = 2\pi \times \SI{1.36}{\mega\hertz}$. An optical qubit defined by the ground state $\ket*{\downarrow} \equiv \ket*{5S_{1/2} , m_J=-1/2}$ and the metastable excited state $\ket*{\uparrow} \equiv \ket*{4D_{5/2} \:, m_J=-5/2}$ provides an isolated two-level system, to which the two motional modes can be coupled. The tunable coupling is introduced by a laser field on the first red sideband (RSB) transitions $\ket*{\downarrow, n} \leftrightarrow \ket*{\uparrow, n-1}$.

We use a single bichromatic laser beam resonant to both RSB transitions to simultaneously couple the two radial modes to the electronic states. Assuming the two modes are equally coupled by the bichromatic beam the interaction Hamiltonian describing this system reads (see Appendix~\ref{appendix:bright and dark states})
\begin{equation}
	\hat{H}_{\mathrm{int}} \approx \hbar g\left(\hat{\sigma}^{+} \hat{a}_{1} + \hat{\sigma}^{-} \hat{a}^{\dagger}_{1}\right)
	+ \hbar g \left(\hat{\sigma}^{+} \hat{a}_{2} + \hat{\sigma}^{-} \hat{a}^{\dagger}_{2}\right),
	 \label{Eq:RBS}
\end{equation}
where $g$ is the coupling strength,  $\hat{a}_{i}$ ($\hat{a}^{\dagger}_{i}$) is the annihilation (creation) operator of the radial mode $i \in \{1,2\}$ and $\hat{\sigma}^{+} \equiv \ket*{\uparrow}\bra*{\downarrow} \equiv (\hat{\sigma}^{-})^{\dagger}$ is the raising electronic operator for the ion. The intensities of the two tones of the bichromatic field are chosen to produce equal coupling strength, resulting in equal Rabi frequencies for the two RSB transitions. The associated Lamb-Dicke parameters were measured to be $\eta_1 = 0.041$ and $\eta_2 = 0.044$ (see Appendix~\ref{appendix:bright and dark states}).

Here we introduce the two-mode $N-$phonon basis states $\ket*{\psi_n^N}$ which are coupled to each other by $\hat{H}_{\mathrm{int}}$ via the following relation 
\begin{equation}
    \hat{H}_\mathrm{int} \ket*{\psi_n^N} \ket*{\downarrow} = g \sqrt{2n} \ket*{\psi_{n-1}^{N-1}} \ket*{\uparrow},
    \label{Eq:basis}
\end{equation}

\noindent where $0\leq n\leq N$ (for a complete description of the states $\ket*{\psi_n^N}$, see Appendix~\ref{appendix:bright and dark states}). We focus on the extremal cases of $n = 0$ and $n = N$, which produce the maximally bright and perfectly dark states
\begin{align}
    \ket*{B^N} &\equiv \ket*{\psi_{N}^{N}} &= \sqrt{\frac{N!}{2^N}}\sum_{m=0}^{N}\frac{1}{\sqrt{m!(N-m)!}}\ket*{m,N-m},\\
    \ket*{D^N} &\equiv\ket*{\psi_{0}^{N}} &= \sqrt{\frac{N!}{2^N}}\sum_{m=0}^{N}\frac{(-1)^m}{\sqrt{m!(N-m)!}}\ket*{m,N-m}.
 \end{align}

\noindent Analogously to a dark state of light, a \textit{non-trivial} dark state $\ket*{D^N}$ is a collective state of both phonon modes satisfying $\hat{H}_{\mathrm{int}}\ket*{\downarrow}\ket*{D^N} = 0$ and $\langle D^N|\hat{N}|D^N\rangle > 0 $, with the total excitation number operator $\hat{N} = \hat{a}_{1}^{\dagger}\hat{a}_{1} + \hat{a}_{2}^{\dagger}\hat{a}_{2}$. The dark state is therefore characterized by a completely suppressed ion-phonon coupling and therefore no excitation of the ion by the incoming light field. A bright state will exhibit the opposite effect, in which the ion-phonon coupling is enhanced and population is transferred at a faster rate than can be achieved for the case of an ion interacting with a single motional mode with the same amount of phonons [described by reducing the Hamiltonian of Eq.~\eqref{Eq:RBS} to include only one of the two terms]. Thus, the state $\ket*{B^N}$ adheres to the definition $\bra*{B^{N-1}}\bra*{\uparrow}\hat{H}_{\mathrm{int}}\ket*{\downarrow}\ket*{B^N}/\hbar = \sqrt{2N}g$. For the case of a single phonon shared between the two motional modes the bright and dark states are 
\begin{align}
	&\ket*{B^{1}}=\frac{1}{\sqrt{2}}\Bigl( \ket*{0,1} + \ket*{1,0} \Bigr) \label{Eq:brightCoupling}\\
 	&\ket*{D^{1}}=\frac{1}{\sqrt{2}} \Bigl( \ket*{0,1} - \ket*{1,0} \Bigr).
  \label{Eq:darkCoupling}
\end{align}

\noindent Note that here and throughout the remainder of the article explicit labels for modes $1$ and $2$ have been omitted for clarity. The single-phonon bright state then couples to the atom as $\bra*{0,0}\bra*{\uparrow}\hat{H}_{\mathrm{int}}\ket*{\downarrow}\ket*{B^{1}}/\hbar = \sqrt{2}g$.

Evidence of dark and bright phononic states can be probed experimentally by monitoring the dynamical population of the electronic state of a single ion evolving under the action of the bichromatic field. The internal state is determined via fluorescence detection from the ground state $\ket*{\downarrow}$ to the short-lived excited states $\ket*{5P_{1/2} \:, m_J=\pm1/2}$. The motional state is measured for each mode individually by driving Rabi oscillations on blue sideband (BSB) and RSB transitions following the approach in Ref.~\cite{Leibfried:03}.


 \section{Interference with a single phonon}
 \label{section:quantum_section}

\begin{figure}[t]
        \includegraphics[width=\columnwidth]{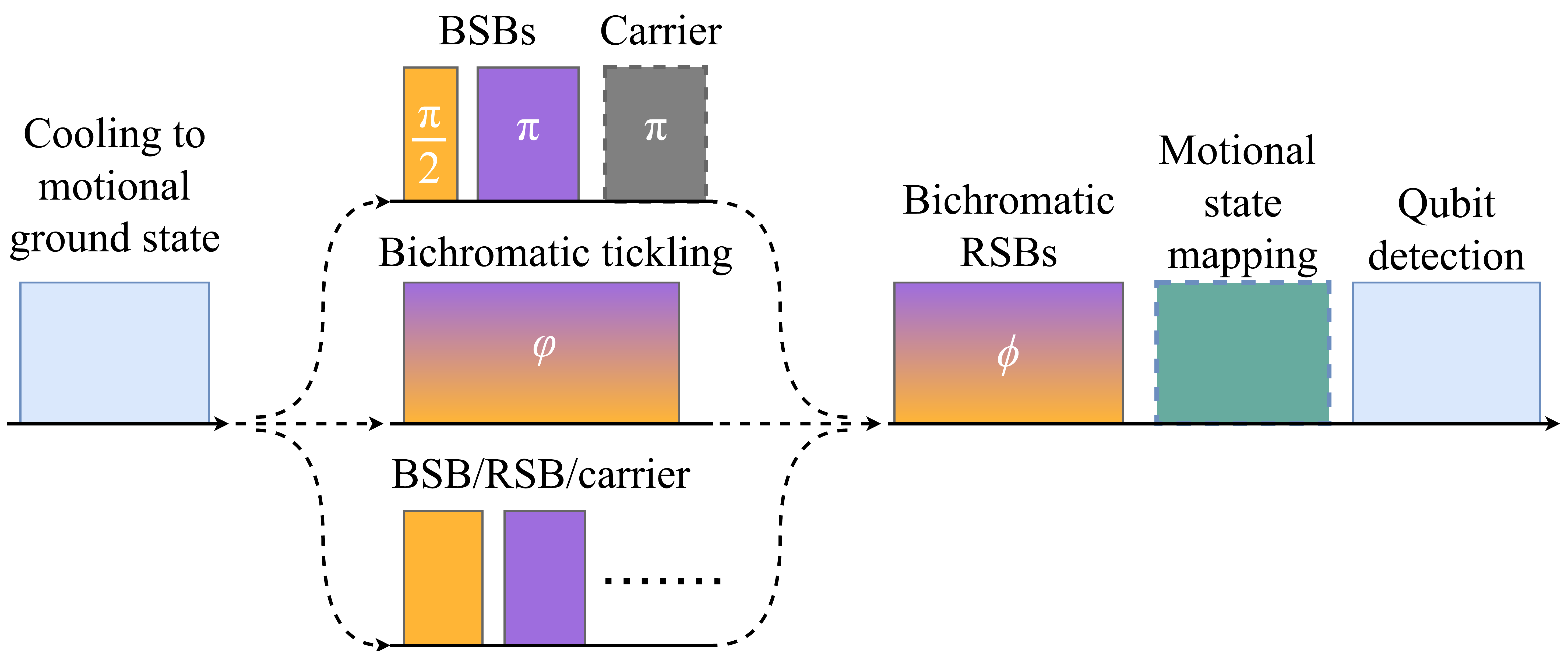}
        \caption{Simplified pulse sequences for the experiments. The basic pulse sequence depicted is the same for all cases investigated in this work, only the initial motional preparation differs. In all cases, the ion is first initialized in the electronic and motional ground state. From top to bottom, the subsequent pulses are used to prepare the single phonon states described in Sec.~\ref{section:quantum_section}, the coherent states described in Sec.~\ref{section:coherent_section} and the single phonon product states described in Sec.~\ref{section:coherent_section} respectively. The state preparation is always followed by a bichromatic pulse of variable length. Finally, the state of the ion is readout by fluorescence detection. In some cases, the motional state of the ion is first mapped to the electronic states by additional RSB and BSB pulses.}
        \label{fig:all_pulse_seqs}
\end{figure}

The pulse sequence used to prepare bright and dark states is shown in Fig.~\ref{fig:all_pulse_seqs}. The initialization procedure consists of Doppler cooling followed by resolved sideband cooling, by which we achieve mean phonon numbers $\tilde{n}_{1,2} < 0.1 $. At this point, the ion is nominally in state $\ket*{\downarrow,0,0}$. For the single phonon states we then perform a $\pi/2$ pulse on the BSB transition of mode $1$ followed by a $\pi$ pulse on the BSB transition of mode $2$, creating the superposition state $\ket*{\uparrow}(\ket*{0,1}+\ket*{1,0})/\sqrt{2}$. We subsequently perform a fluorescence measurement and discard all instances where the population transfer to $\ket*{\uparrow}$ was not successful. The ion is then brought back to the electronic ground state $\ket*{\downarrow}$ by a carrier $\pi$ pulse (see Appendix.~\ref{appendix:single phonon prep} for further details).

\begin{figure}[hb]
    \includegraphics[width=0.9\columnwidth]{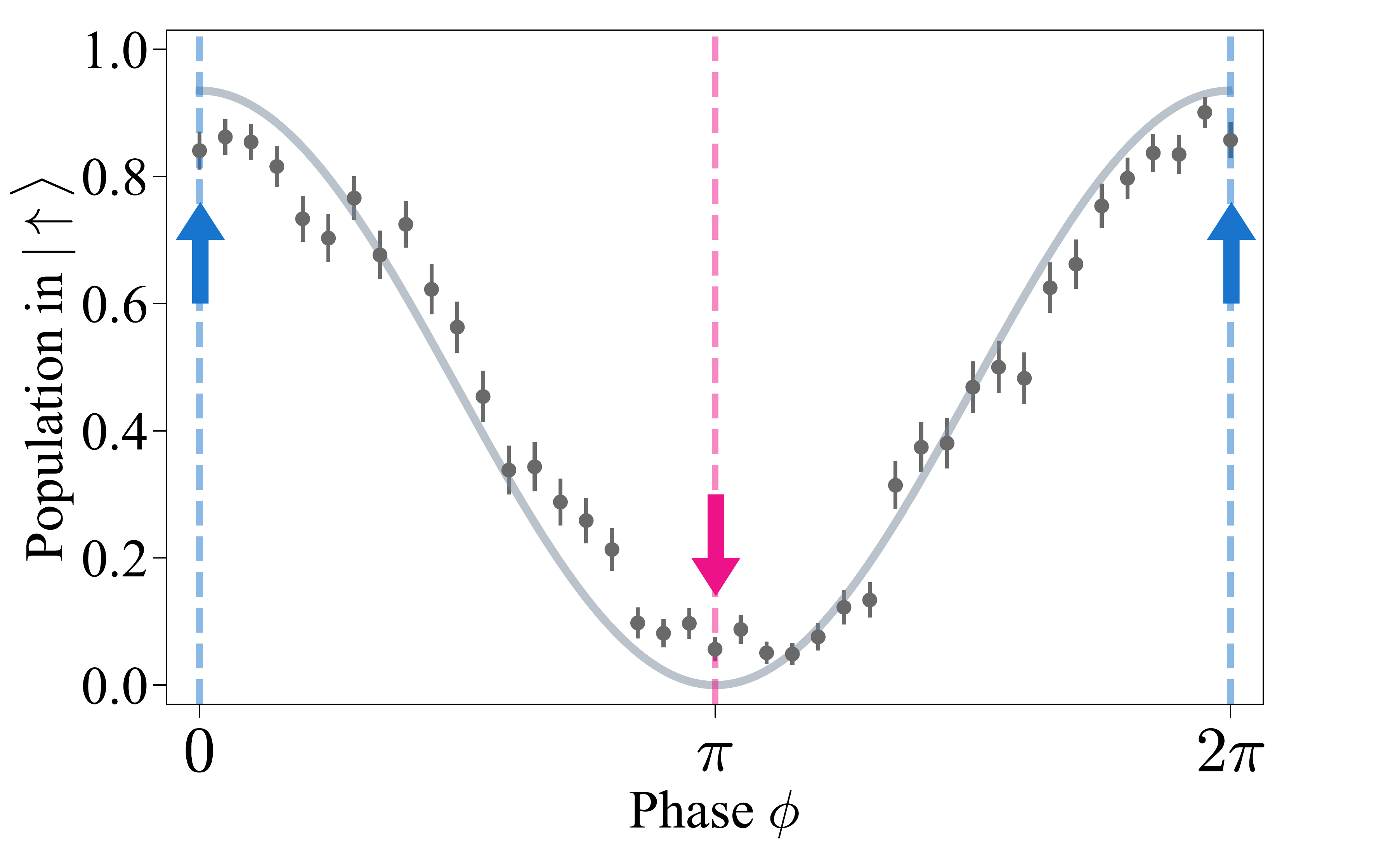}

    \vspace{0.5em}

    \centering
   \includegraphics[width=0.9\columnwidth]{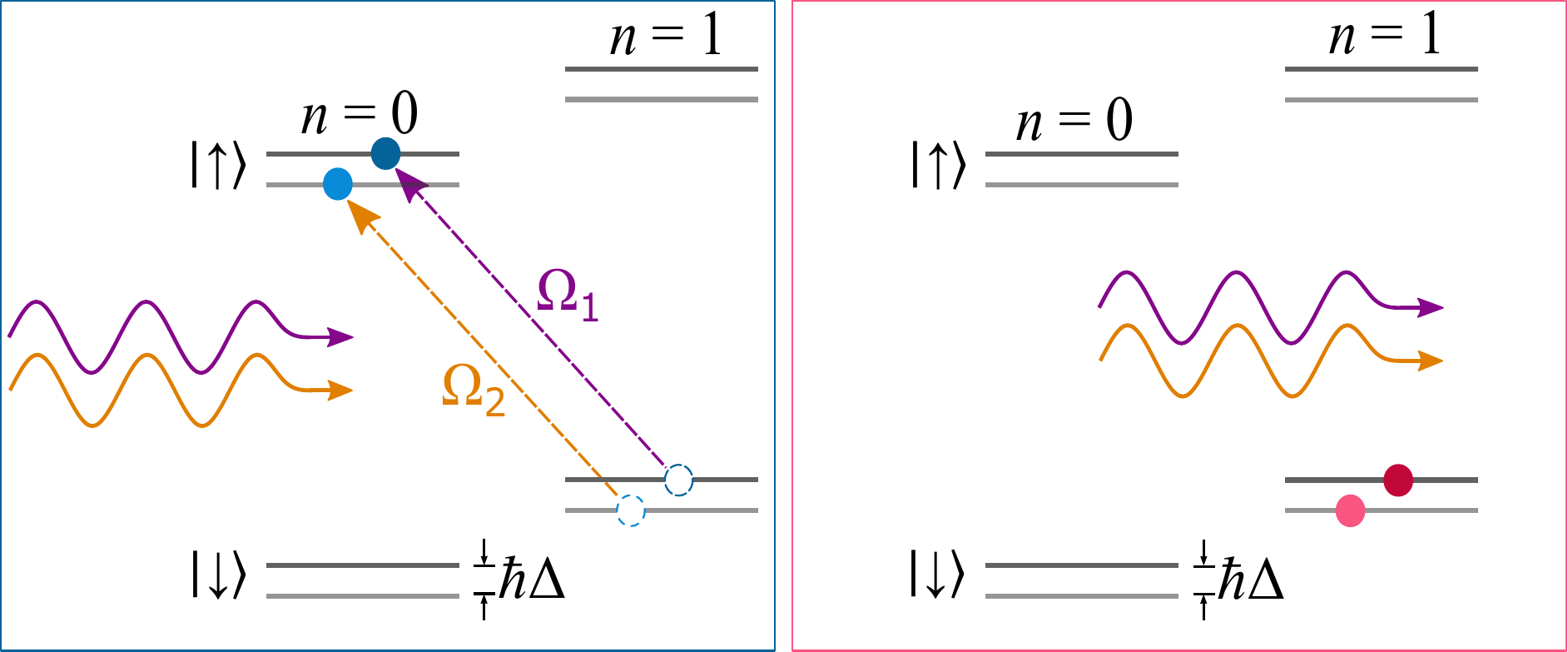}
    \begin{picture}(0,0) 
        \put(-225,223){(a)}       
    \end{picture}
    \begin{picture}(0,0) 
        \put(-224,83){(b)}       
    \end{picture}
    \begin{picture}(0,0) 
        \put(-113,83){(c)}       
    \end{picture}
	\caption[width=\textwidth]{(a) Phase evolution of a two-mode state between bright (blue arrows) and dark (red arrow) coupling regimes when a single quantum is present in the system. (b) and (c) display the relevant transitions for the bright and dark coupling regimes respectively; coloured circles indicate which states are populated after the application of a bichromatic pulse. The bright state sees the incoming light field and population is transferred whereas the dark state remains unaffected.}
    \label{fig:single_phonon_phase}
\end{figure}
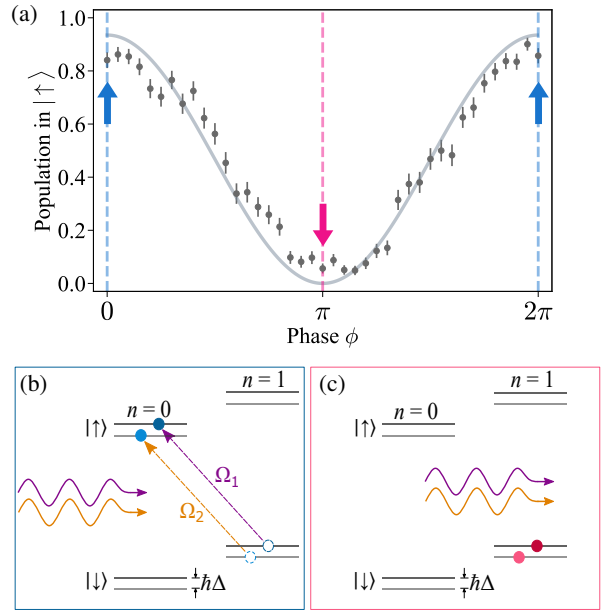

From there we perform the bichromatic RSB pulse. We change the relative phase $\phi$ between the two laser fields instead of changing the phase of the motional state; this introduces a phase for the second term of the interaction Hamiltonian $\hat{H}_{\mathrm{int}}$, resulting in the same dynamics. Since $\phi$ is set at the beginning of each measurement sequence and for each mode the preparation and bichromatic pulse component maintain a stable phase relationship, we can treat the initial state as $(\ket*{0,1}+e^{i\phi}\ket*{1,0})/\sqrt{2}$ and refer to $\phi$ as the relative phase between the two modes. The state at the end of the state preparation with $\phi=0$ corresponds to the expected bright state given by Eq.~\eqref{Eq:brightCoupling}. In Fig.~\ref{fig:single_phonon_phase}(a) we demonstrate that tuning $\phi$ during the application of the bichromatic pulse leads to oscillations in the excited state population. The maxima and minima of this curve are identified as being the bright and dark coupling regimes respectively. Fig.~\ref{fig:single_phonon_phase}(b) shows the transitions involved in the transfer from the bright state. Fig.~\ref{fig:single_phonon_phase}(c) highlights the lack of transfer from the dark state. The solid line in Fig.~\ref{fig:single_phonon_phase}(a) is the theoretical evolution of the state in which $\hat{H}_{\textrm{int}}$ is applied for a fixed time, based on the coupling strength $g$. The coupling strength $g$ was determined from separate measurements of the RSB Rabi frequencies $\Omega_1$ and $\Omega_2$. 

\begin{figure*}[t]
    \begin{minipage}{\columnwidth}
        \vspace{-1em}
        \hspace*{-1.5em}
        \includegraphics[width=1.17\textwidth]{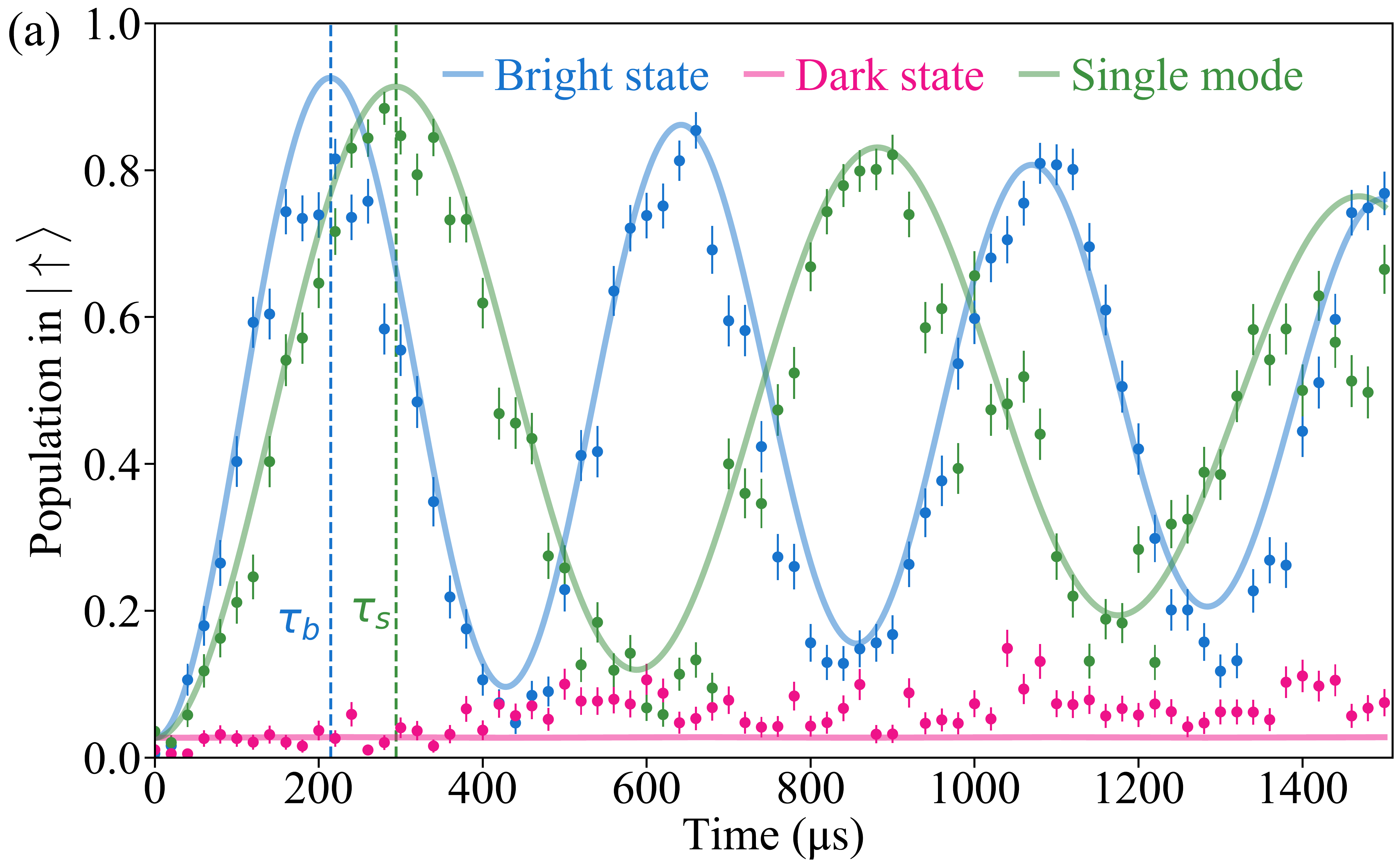}
        \vspace{-0.75em}
        \hspace{1.5em}
    \end{minipage}
    \begin{minipage}{\columnwidth}
        \begin{flushright}
        \hspace*{3.5em}
        \includegraphics[width = 0.9\textwidth]{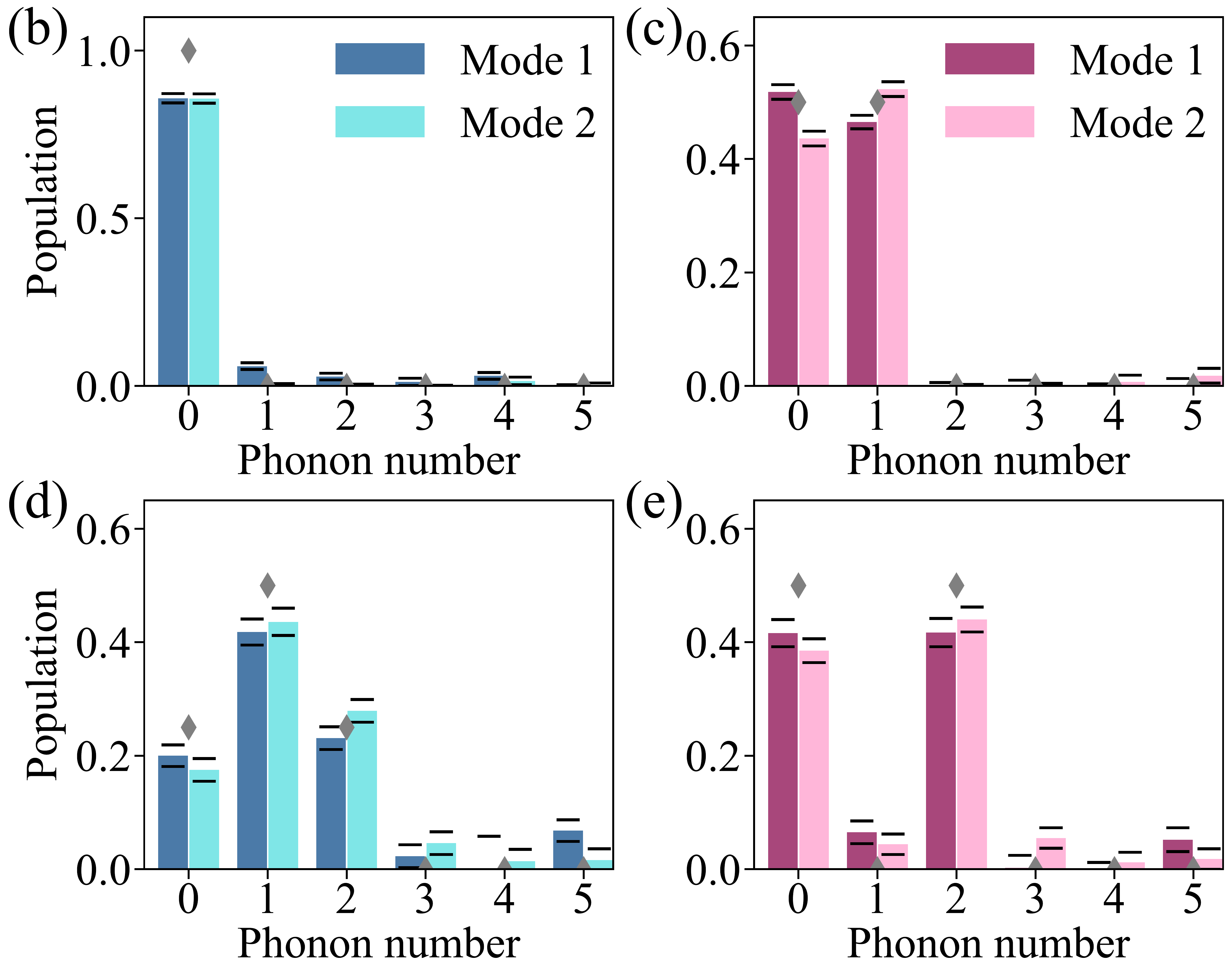}
        \begin{picture}(0,0)
            \put(-15,136){\includegraphics[height=0.6cm]{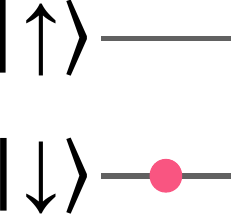}}
        \end{picture}
        \begin{picture}(0,0)
            \put(-128,136){\includegraphics[height=0.6cm]{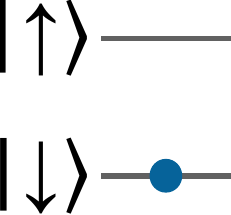}}
        \end{picture}
        \begin{picture}(0,0)
            \put(-17,72){\includegraphics[height=0.6cm]{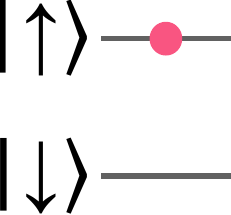}}
        \end{picture}
        \begin{picture}(0,0)
            \put(-131,72){\includegraphics[height=0.6cm]{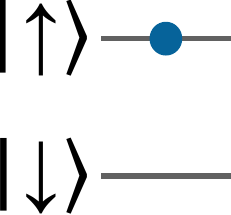}}
        \end{picture}
        \end{flushright}
    \end{minipage}
    \vspace{-1em}
    \caption[width=\textwidth]{(a) An ion in its electronic ground state $\ket*{\downarrow}$ evolves differently from the action of $\hat{U}_{\textrm{int}}$ depending on the preparation of its motional state. The Rabi frequency of the bright state ($\ket*{B^1}$) is enhanced compared to the single-mode-excitation state ($\ket*{0,1}$), conversely Rabi oscillations for the dark state ($\ket*{D^1}$) are largely suppressed. In the latter case, residual coupling to $\ket*{\uparrow}$ is due to the limited efficiency of the bichromatic pulse and decoherence. The solid lines show the simulated evolution of each state with optimized experimental parameters. (b-c) Phonon number distributions after a bichromatic pulse of length $\tau_b$ is applied for the bright and dark coupling regimes respectively. Gray diamonds indicate the theoretically ideal populations. The bright state couples to the available modes and the single phonon is removed from the system. The dark state is not coupled to any available mode and the single phonon superposition state is retained. (d-e) Phonon number distributions after the bichromatic pulse is applied to an ion in the excited state $\ket*{\uparrow}$. In (d) an additional phonon is added during the population transfer but without leaving the bright state subspace. In (e) the excited ion interacts with the dark state just as it would for a single vacuum mode, in this case, the addition of a phonon causes the ion to leave the dark state subspace. Error bars indicate $1\sigma$ confidence intervals for quantum projection noise in (a) and fitting error in (b-e).}
   \label{fig:ground_evo} 
\end{figure*}

We note that the efficiency of the bichromatic pulse is impacted by coupling to the carrier transition (both resonantly due to laser frequency noise in the MHz range and off-resonantly via the AC Stark effect) and by small differences in the intensity of the two frequency components. These technical limitations could be improved by reducing servo-induced frequency noise in the laser output \cite{Lintao:01}, increasing the available laser power to allow for the implementation of AC Stark shift compensation \cite{Haffner:01} and by implementing finer amplitude control of the two modulation signals. 

The bichromatic pulse acts as $\hat{U}_{\textrm{int}} = e^{i\hat{H}t/\hbar}$ where $\hat{H}$ is the Hamiltonian described by Eq.~\eqref{Eq:RBS}. With $\phi = 0$ the evolution of the state is driven as $\smash{\ket*{\downarrow}\ket*{B^1} \myrightxarrow{\hat{U}_{\textrm{int}}} \ket*{\uparrow} \ket*{0,0}}$.
We demonstrate the effect of preparing a bright state on the ion-phonon coupling by driving Rabi oscillations and observing a $\sqrt{2}$ enhancement for the bright state compared with the single-mode state $\ket*{\downarrow,0,1}$, where no collective behaviour is expected. The results are shown in Fig.~\ref{fig:ground_evo}(a). Here the solid lines are again produced by a simulation based on the Hamiltonian from Eq.~\eqref{Eq:RBS}, but in this case the coupling $g$ is a free fit parameter and the electronic and motional decoherences are taken into account. Final fit values for these parameters are consistent with experimental expectations (see Appendix~\ref{appendix:simulations} for more details on the theoretical model used for this simulation and those of subsequent figures in the main text). To verify that the enhancement we observe is due to the interference between the two modes and not, for example, off-resonant driving of the carrier transition, we also measure the phonon number distribution of each mode individually after a bichromatic pulse of length $\tau_b$  has been applied; $\tau_b$ corresponds to the maximum population transfer observed for the blue curve in Fig.~\ref{fig:ground_evo}(a). The results in Fig.~\ref{fig:ground_evo}(b) show that both modes are returned to their respective ground states with high probability. This indicates that the electronic state population transfer occurs predominantly due to the bright state coupling.

With $\phi = \pi$ the system is ideally in a perfectly dark state and destructive interference between the available paths $\ket*{\downarrow,1,0} \rightarrow \ket*{\uparrow,0,0}$ and $\ket*{\downarrow,0,1} \rightarrow \ket*{\uparrow,0,0}$ leads to a complete suppression of the ion-phonon coupling, even though phonons are present in the system. Fig.~\ref{fig:ground_evo}(a) shows that the dark state suppression inhibits any significant population transfer such that we are completely unable to drive Rabi oscillations. Similarly, in Fig.~\ref{fig:ground_evo}(c) almost no change in the phonon number distribution of each mode can be detected after the bichromatic pulse has been applied for a time $\tau_{b}$. The results shown here demonstrate that for a highly non-classical initial state, the atom-boson coupling can be explained in full by considering representations in the collective basis of Eq.~\eqref{Eq:basis}. Maximal coupling, which is a consequence of constructive interference between the available modes, is defined by a decomposition solely onto the bright state subspace of this basis [Eq.~\eqref{Eq:brightCoupling}]. On the other hand, completely destructive interference is defined by a decomposition solely onto the dark state subspace [Eq.~\eqref{Eq:darkCoupling}]. For the single-mode-excitation state $\ket*{0,1}$, an absence of interference effects can be explained by its decomposition onto the equal superposition $(\ket*{B^1}+\ket*{D^1})/\smash{\sqrt{2}}$. 

By removing the final carrier $\pi$ pulse from the preparation stage of the experiment, it is also possible to study how the interaction between the two modes and the single ion is altered when the ion is not in its electronic ground state. Here we have to consider multiple RSB transitions $\ket*{\uparrow,n-1, m-1} \leftrightarrow \ket*{\downarrow,n, m}$ with coupling strengths that scale with the phonon number as $\sqrt{n\cdot m}$. With the final carrier pulse removed, the case of $\phi = \pi$ is drastically altered as $\hat{H}_{\mathrm{int}}\ket*{\uparrow}\ket*{D^1} \neq 0$. With the ion in an excited state only two of the four available paths will destructively interfere, that is the transitions $\ket*{\uparrow,1,0} \rightarrow \ket*{\downarrow,1,1}$ and $\ket*{\uparrow,0,1} \rightarrow \ket*{\downarrow,1,1}$ will cancel one another but the transitions $\ket*{\uparrow,1,0} \rightarrow \ket*{\downarrow,2,0}$ and $\ket*{\uparrow,0,1} \rightarrow \ket*{\downarrow,0,2}$ will not. 

By choosing appropriate lengths for the bichromatic pulse in each case ($\phi = 0$ and $\phi = \pi$, see Appendix~\ref{appendix:excited state single phonon} for details), we again measure the phonon number distributions of the individual modes after a $\pi$ pulse has been applied, the results are given in Figs.~\ref{fig:ground_evo}(d) and ~\ref{fig:ground_evo}(e). For $\phi = 0$ the measured populations resemble those of the two-phonon bright state \cite{Maximo:01}, $\ket*{B^2} = \left(\ket*{0,2}+\smash{\sqrt{2}}\ket*{1,1}+\ket*{2,0}\right)/2$. Thus for an atom in an excited state $\hat{U}_\mathrm{{int}}$ acts analogously to a raising ladder operator in the Fock basis; climbing higher within the bright state manifold is therefore possible by the repeated application of bichromatic and carrier pulses. For $\phi = \pi$ the population of each mode is approximately equally distributed between the states with $n=0$ and $n=2$ in a way that is consistent with the NOON state $\left(\ket*{0,2}+\ket*{2,0}\right)/\smash{\sqrt{2}}$. In this case we leave the dark state manifold and the action of $\hat{U}_{\mathrm{int}}$ can be understood as a phononic beamsplitter operation, with the two output ports corresponding to the two available modes.


\section{Interference with coherent states}
 \label{section:coherent_section}
 
In this section, we demonstrate the validity of the bright and dark state formalism in describing \textit{classical-like} interference effects by preparing a set of two-mode coherent states of the form
\begin{equation}
    \ket*{\Psi_{\alpha, \varphi}} = \ket*{\alpha, e^{i \varphi}\alpha}.
    \label{Eq:coherent}
\end{equation}

\noindent The phase $\varphi$ determines whether the two fields will constructively add or destructively cancel. The former case, with $\varphi = 0$ corresponds to the two-mode bright state \cite{Maximo:01} 
\begin{equation}
\ket*{B^{\alpha}} \equiv \ket*{\alpha, \alpha} = e^{-\left|\alpha\right|^2}\sum_{N=0}^\infty\sqrt{\frac{2^N}{N!}}\alpha^N\ket*{B^N}.
\end{equation}
The latter case, with $\varphi = \pi$ corresponds to the two-mode dark state
\begin{equation}
\ket*{D^{\alpha}} \equiv \ket*{\alpha, -\alpha} = e^{-\left|\alpha\right|^2}\sum_{N=0}^\infty\sqrt{\frac{2^N}{N!}}\alpha^N\ket*{D^N}.
\end{equation}
The interference effects are therefore characterized in exactly the same manner as the highly non-classical states in Sec.~\ref{section:quantum_section} --- a decomposition of the initial state onto a subspace that is made up of solely bright or dark collective states.

Coherent states are prepared by tickling the ion with an electric field oscillating at the trapping frequency. The field is created by a modulation of the electric potential on an electrode close to the ion. Ideally the field produced by this electrode has a radial component that overlaps equally with the two motional modes, however imperfections on the electrodes surface and stray charges can cause a slight discrepancy in the rates of excitation (see Appendix~\ref{appendix:coherent calibration} for a calibration of the response of each mode to the driving field). We can precisely control the coherent excitation by varying the duration, amplitude and phase of the tickling. To create a two-mode coherent state, we modulate the electrode with both trapping frequencies at the same time with a fixed phase $\varphi$ between the signals. To ensure phase coherence between the state creation and the application of $\hat{U}_{\textrm{int}}$ we use the same frequency sources to tickle the electrodes and to drive the acousto-optic modulator which produces the bichromatic pulse.

 \begin{figure}[bp]
    \begin{center}
   \includegraphics[width=\columnwidth]{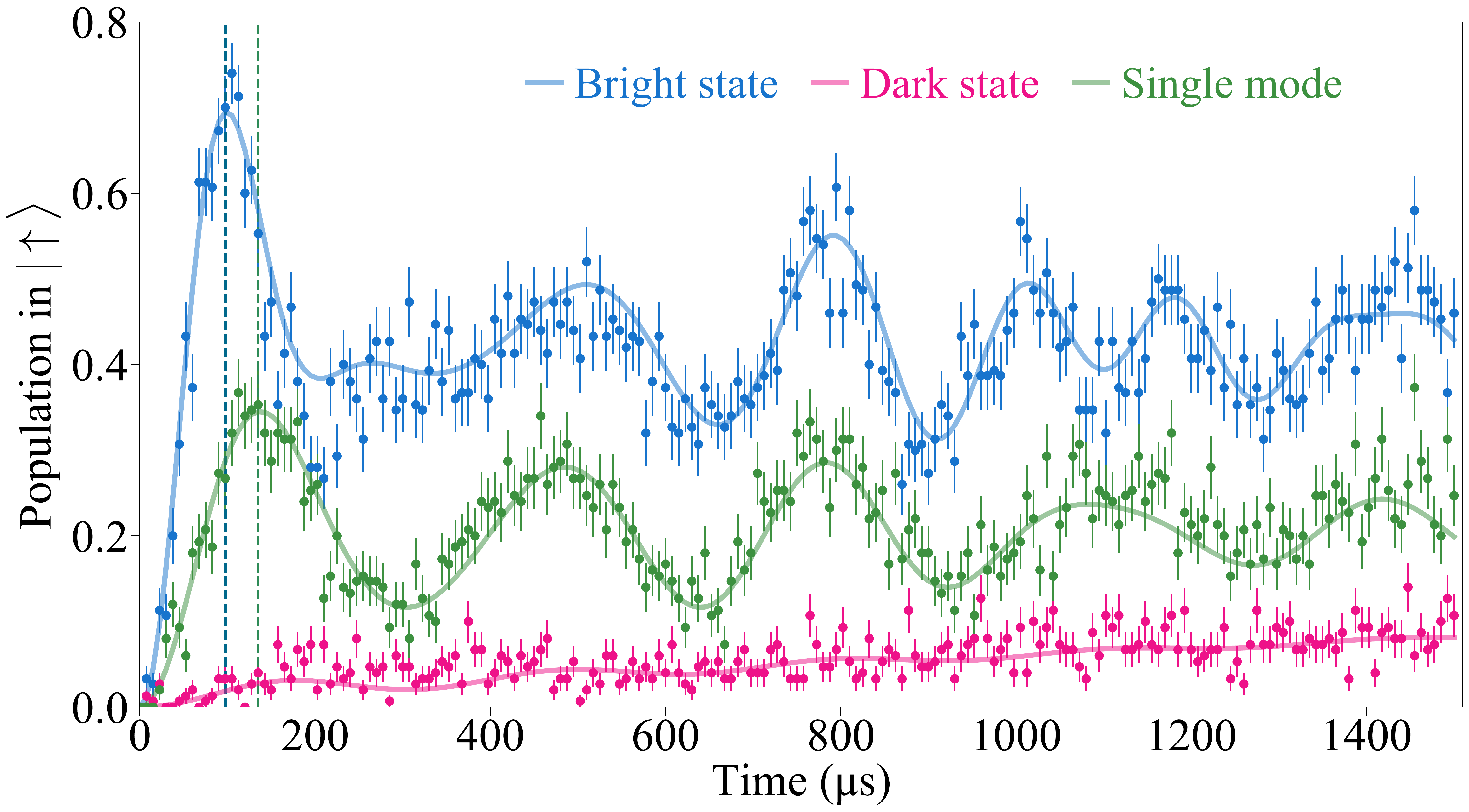}           
    \end{center}
    \vspace{-1.5em}
    \caption[width=1.3\textwidth]{Population transfer under the action of $\hat{U}_{\textrm{int}}$ for two-mode coherent states $\ket*{\downarrow, \alpha, \alpha}$ (blue), $\ket*{\downarrow, \alpha, -\alpha}$ (pink) and $\ket*{\downarrow, \alpha, 0}$ (green) with $\alpha = 1$. An ion initially prepared in $\ket*{\downarrow}$ experiences an enhanced/suppressed coupling for the coherent bright/dark states.
    Residual excitation to $\ket*{\uparrow}$ from the dark state is due to the finite temperature after sideband cooling and decoherences, however no oscillatory behaviour is observed. Decoherence is primarily attributed to laser phase fluctuations, motional dephasing is comparatively negligible (see Appendix~\ref{appendix:simulations}). Error bars depict the $1\sigma$ quantum projection noise, the solid lines show the theoretically expected evolution.}
    \label{fig:coherent_evo}
\end{figure}

\begin{figure*}[t!]
    \begin{flushleft}
        \begin{picture}(0,0) 
            \put(-8,150){(a)}       
        \end{picture}
        \raisebox{0em}{\includegraphics[width=1.1\columnwidth]
        {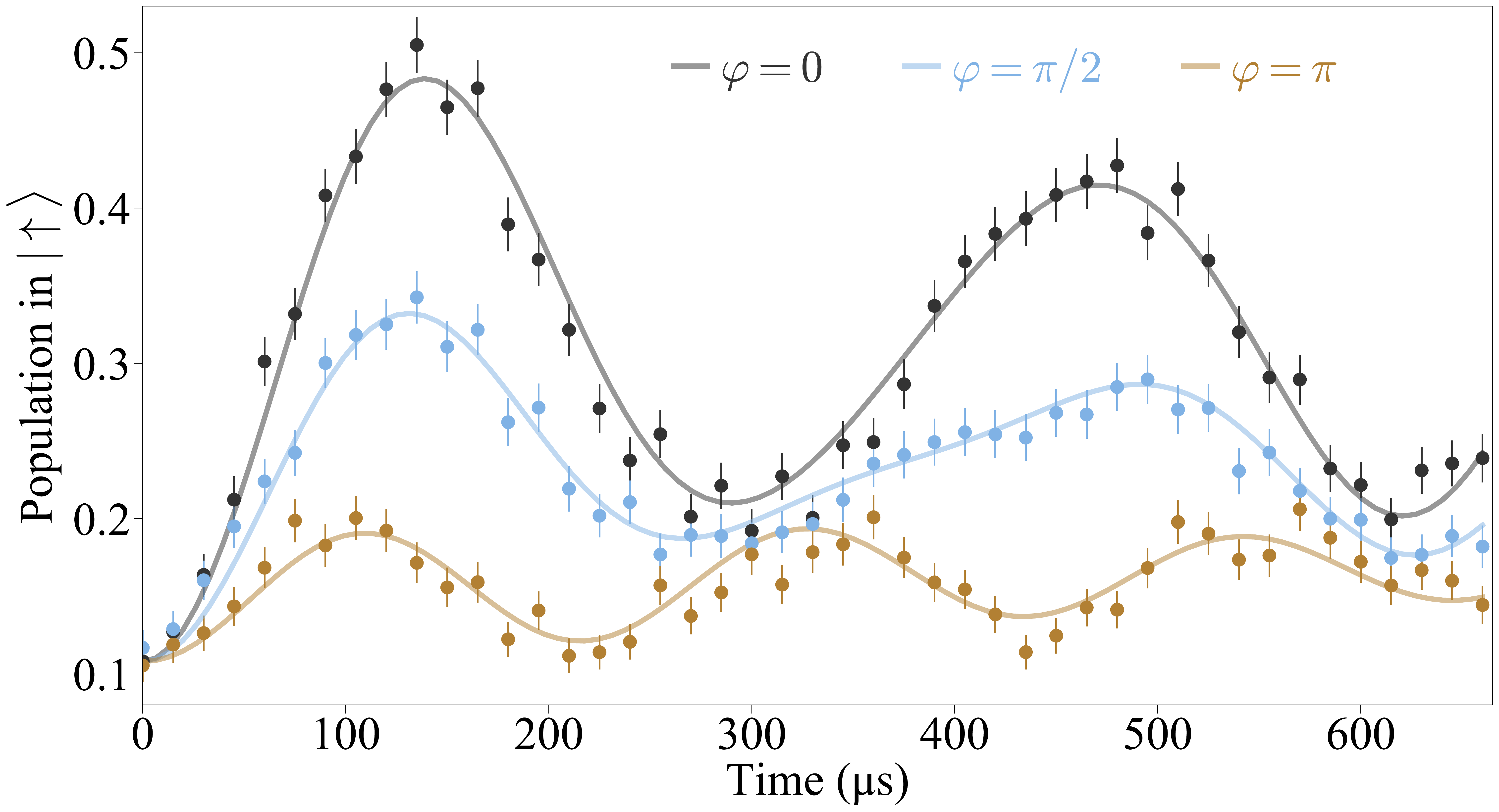}}
        \hspace{2em}
        \begin{picture}(0,0) 
            \put(-8,150){(b)}       
        \end{picture}
        \includegraphics[width=0.7\columnwidth]{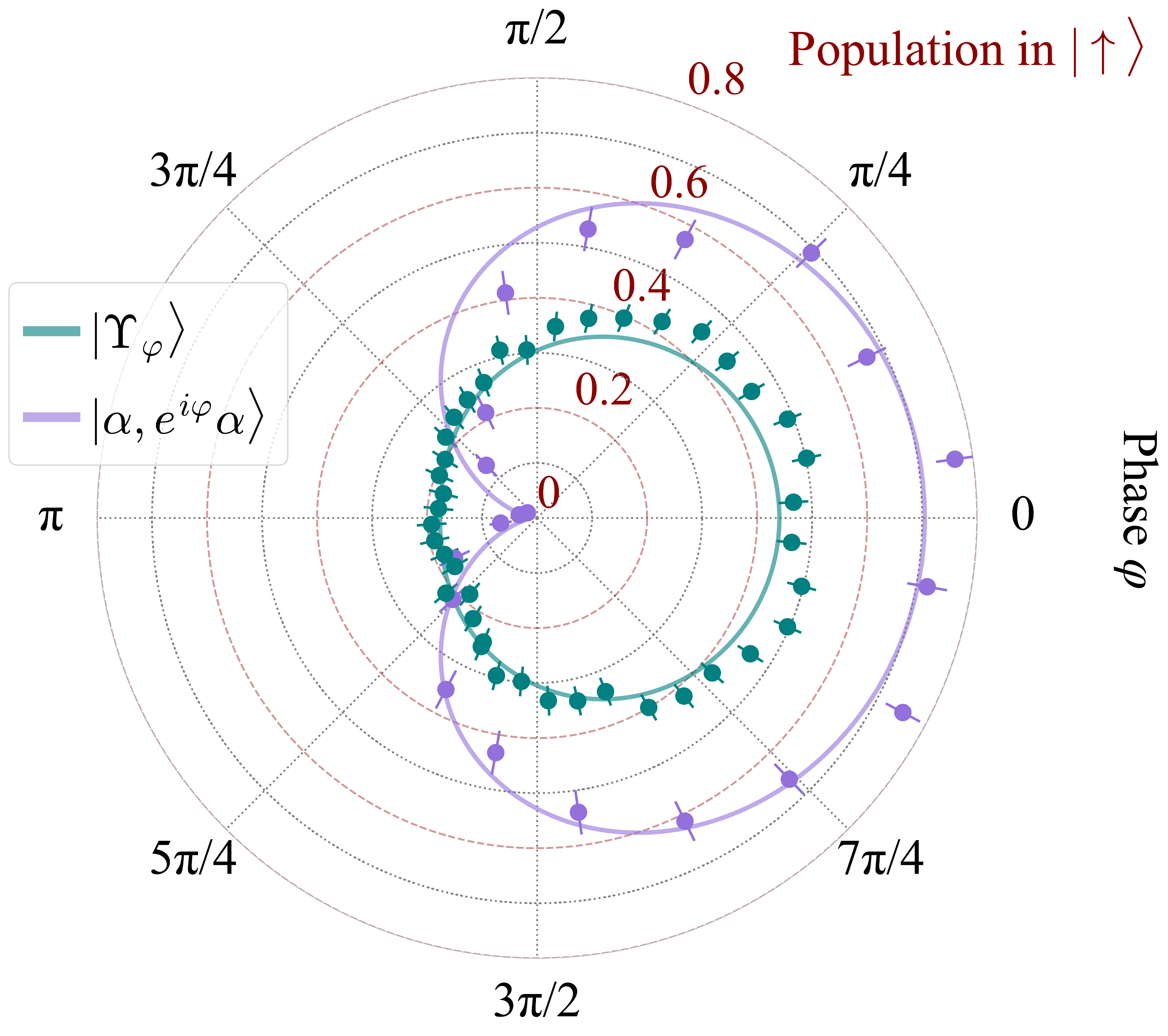}
        \begin{picture}(0,0) 
            \put(-21,62){\includegraphics[height = 0.12\columnwidth]{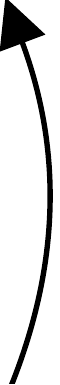}}
        \end{picture}
    \end{flushleft}

    \vspace{-1em}
    
        \caption{(a) Evolution of the state $\ket*{\Upsilon_{\varphi}}$ under the action of $\hat{U}_{\textrm{int}}$. For all three phase values ($\varphi = 0$, $\varphi = \pi/2$, and $\varphi = \pi$), there is strong agreement between the experimental data (dots) and the simulated evolution (solid lines). Crucially, low-amplitude Rabi oscillations for the state $\ket*{\Upsilon_{\pi}}$ indicate a non-zero coupling between the electronic state and the motional modes. This coupling is neatly explained by the contribution of a two-phonon bright state in Eq.~\eqref{eq:upsilon_collective}. (b) Phase scans for the Upsilon state $\ket*{\Upsilon_\varphi}$ and the two-mode coherent state $\ket*{\alpha, e^{i\varphi}\alpha}$, after a bichromatic $\pi$ pulse has been applied in each case. Again, there is reasonable agreement between the experimental data points (dots) and the simulated results (solid lines). For the two-mode coherent state, the excitation drops to zero for a phase of $\varphi = \pi$, but not for the $\Upsilon$ state despite both sharing the same average field strength and variance. Error bars indicate quantum projection noise ($1\sigma$ confidence intervals).}
        \label{fig:upsilon_evolution}
\end{figure*}
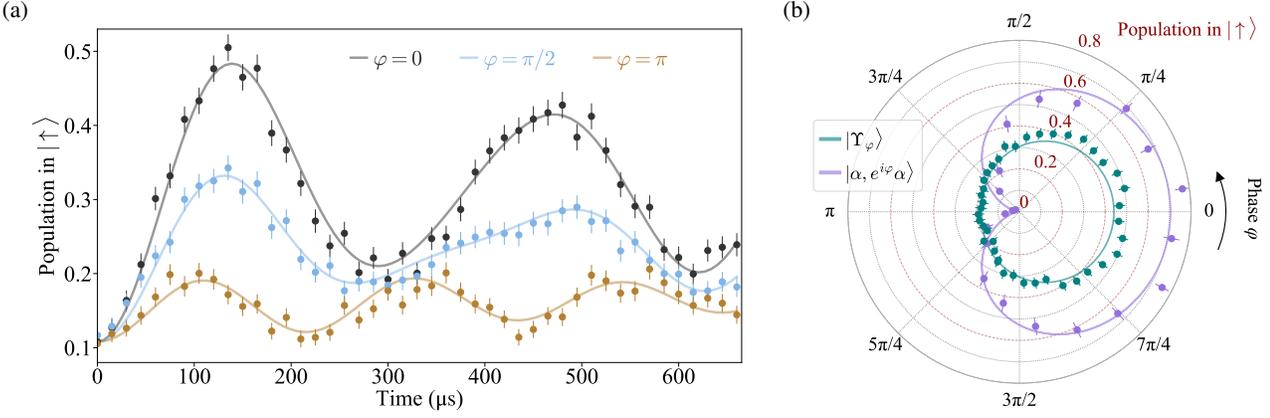

To see the expected constructive and destructive interference, we probe the evolution of the bright, dark, and single-mode-excitation ($\ket*{\alpha, 0}$) states under the action of $\hat{U}_{\textrm{int}}$. For each of the three initial states we perform tickling to produce a coherent excitation of $\alpha = 1$. Although the average phonon number in this case remains low ($\Bar{n} = 1$) the spread of $n$ adheres to a Poissonian distribution and therefore allows us to probe \textit{classical-like} interference effects whilst remaining definitively within the Lamb-Dicke regime (see Appendix~\ref{appendix:coherent calibration} for further details). The results are shown in Fig.~\ref{fig:coherent_evo}; enhanced bright state coupling is evidenced by the increase in Rabi frequency as compared to the single-mode-excitation state. The overall higher population transfer of the bright state is due to the fact that both modes are coupled to the ion's electronic state; for the single-mode state roughly $50\%$ of the initial population resides in the motional ground state, to which the RSB transitions do not couple. For the dark state, we see a strong suppression of the coupling as is expected. The residual population transfer can be explained by a finite temperature after cooling the ion, heating effects in the state preparation, and both electronic and motional dephasing. 

A simulation of the dynamics is shown by the solid lines and is in very good agreement with the experimental data. All simulated curves presented in Fig.~\ref{fig:coherent_evo} share the same coupling strength and decoherence parameters. An initial state of
\begin{equation}
    \rho_{\text{init}} = \hat{D}(\alpha) \rho_{\textrm{th}} \hat{D}(\alpha)^\dagger
\end{equation}
was used with displacement operators $\hat{D}(\alpha)$ with $\alpha = 1$ and thermal density matrix $\rho_{\textrm{th}}$ with $n_{\textrm{th}}=0.025$. The temperature after ground state cooling and coherent displacement parameters were determined in separate experiments. 

For coherent states of light, a non-zero atom-light coupling is often characterized in terms of the average electric field $\langle E \rangle$ at the atom's position. This ``classical'' requirement of the light field holds true for the results presented in Fig~\ref{fig:coherent_evo}. For highly non-classical Fock states such as those prepared in Sec.~\ref{section:quantum_section}, which simulate a light field with $\langle E \rangle = 0$, a non-zero atom-light coupling would most often be characterized by a non-zero variance of the electric field $\langle \Delta E \rangle^2 \neq 0$. Thus far, the results predicted by the newly introduced bright and dark state formalism have not deviated from either of these ``typical'' requirements. To highlight that the bright and dark state formalism holds beyond these requirements we also investigate the coupling of the product state
\begin{equation}
    \ket*{\Upsilon_\varphi} = \frac{1}{2} \Bigl( \ket*{0} + \ket*{1} \Bigr) \left( \ket*{0} + e^{i \varphi} \ket*{1} \right).
    \label{Eq:upsilon}
\end{equation}
Interestingly, with $\varphi = \pi$, the properties $\langle E \rangle = 0$ and $\langle \Delta E \rangle^2 = 2$ are shared by both $\ket*{\Upsilon_{\pi}}$ and the coherent dark state $\ket*{D^{\alpha}}$. Here the advantage of the bright and dark state formalism becomes apparent: according to the typical requirements for atom-light interaction, $\ket*{\Upsilon_{\pi}}$ and $\ket*{D^{\alpha}}$ should provide the exact same response to the bichromatic field, namely they should both be dark states of the system. However, $\hat{H}_{\textrm{int}}\ket*{\Upsilon_{\pi}} \neq 0$ and the surviving interaction is left without a viable explanation. Comparatively, in our collective basis $\ket*{\Upsilon_{\pi}}$ takes the form
\begin{equation}
    \ket*{\Upsilon_{\pi}} = \frac{1}{2}\Bigl\{\ket*{D^0}-\sqrt{2}\ket*{D^1}+\frac{1}{\sqrt{2}} ( \ket*{D^2}-\ket*{B^2} ) \Bigr\},
    \label{eq:upsilon_collective}
\end{equation}
and a non-zero coupling between the atom and field can be attributed to the last term --- a contribution from the bright state with $N = 2$.  

To prepare the state $\ket*{\Upsilon_\varphi}$ we use a sequence of multiple sideband and carrier pulses with variable phases and pulse lengths. A detailed description of the state preparation is given in Appendix~\ref{appendix:upsilon state formation}. Coupling to the bichromatic field is experimentally investigated for values of $\varphi = 0$ , $\varphi = \pi/2$ and $\varphi = \pi$, the results are shown in Fig~\ref{fig:upsilon_evolution}(a). Crucially, we observe a non-negligible oscillation in the excited state population when the bichromatic field interacts with the state $\ket*{\Upsilon_\pi}$. For all three experimental data sets the amplitude and oscillation rate of the resonant coupling are in very good agreement with the simulation. Again, the decoherences and amplitude of the simulation were chosen to best match the experimental data. The decoherence and overall amplitude is worse when compared with the two-mode coherent state because of the more complicated state preparation and limited fidelities for each carrier and sideband pulse. Not only does the contribution of a two-phonon bright state in $\ket*{\Upsilon_\pi}$ successfully explain the observed atom-boson coupling, if we compare Eq.~\ref{eq:upsilon_collective} with the collective state decomposition of $\ket*{\Upsilon_0}$, it also provides an intuitive understanding of the enhanced oscillation rate that is observed. Written in the collective basis,
\begin{equation}
    \ket*{\Upsilon_{0}} = \frac{1}{2}\Bigl\{\ket*{D^0}+\sqrt{2}\ket*{B^1}+\frac{1}{\sqrt{2}} ( \ket*{B^2}-\ket*{D^2} ) \Bigr\}.
    \label{eq:upsilon_zero_collective}
\end{equation}
Although the weight of $\ket*{B^2}$ is the same here as in Eq.~\ref{eq:upsilon_collective}, the dominant coupling term in the equation above is the more-heavily weighted $N=1$ bright state $\left(\ket*{B^1}\right)$, for which the overall coupling strength is lower. The consequence of this is a slower, but greater population transfer between the electronic states.

For a final comparison we plot the phase dependence of Eq.~\eqref{Eq:upsilon} during the application of a bichromatic pulse of fixed duration, corresponding to the maximum population transfer observed for $\ket*{\Upsilon_\pi}$ in Fig~\ref{fig:upsilon_evolution}(a). We compare with the phase dependence of Eq.~\eqref{Eq:coherent} plotted for a bichromatic pulse duration that corresponds to the maximum population transfer observed for $\ket*{B^\alpha}$ in Fig~\ref{fig:coherent_evo}. The results are plotted in Fig.~\ref{fig:upsilon_evolution}(b); distance from the origin corresponds to the probability of excitation from $\ket*{\downarrow} \rightarrow \ket*{\uparrow}$. We highlight here that in each case the phase $\varphi = \varphi' - \varphi_0$ where $\varphi'$ is the tunable experimental parameter and $\varphi_0$ is a systematic phase shift that arises during the state preparation. $\varphi_0$ differs between the two preparation sequences; the coherent state formation is significantly faster and the systematic phase offset that we account for here is therefore much smaller (see Appendix~\ref{appendix:upsilon state formation} for a more detailed discussion on the source of $\varphi_0$). For the coherent superposition a true dark state is observed for a phase of $\pi$, accounting for residual excitation due to technical imperfections. The state $\ket*{\Upsilon_\pi}$ however, retains a limited coupling to the excited state.


\section{Conclusion}
In this work, we experimentally verified the existence of phononic bright and dark states for a single trapped ion. We demonstrated that the atom-boson coupling in two-mode systems can be enhanced or suppressed compared to the single-mode case and that such effects can be described in the same manner for both highly non-classical Fock states and coherent states. The results presented here provide strong support for the use of a bright and dark state formalism to effectively explain light-matter coupling in any equivalent system. In particular, we have shown that this formalism is applicable where more conventional requirements for the interaction are no longer suitable. As mentioned in \cite{Maximo:01}, the use of this bright and dark state formalism to describe interference at the quantum level could allow for novel research in the field of quantum optics. Given that recent experiments involving trapped ions \cite{Chen:01} as well as superconducting qubits \cite{Qiao:01} have shown that phononic networks are a potentially viable path towards scalable quantum information processing, the results presented here could also be implemented in the development of linear mechanical quantum computing (LMQC). As an alternative to optical networks, for which quantum advantage has already been demonstrated \cite{Zhong:01, Zhong:02, Madsen:01}, LMQC systems feature deterministic state preparation and high detection efficiencies. Universality by the use of nonlinear gates has also been demonstrated in a hybrid discrete- and continuous-variable system using a single trapped ion \cite{Gan:01}. As well as the formation of a two-phonon NOON state and the generation of bright states with enhanced ion-phonon coupling strengths, creating a single-phonon dark state provides a method to perform conditional multi-qubit operations for hybrid systems \cite{Maximo:01}.   \\


\section*{Acknowledgements}
This work was supported by the Knut \& Alice Wallenberg Foundation (Wallenberg Centre for Quantum Technology [WACQT]), by the Swedish Research Council (Grant No. 2017-04638, 2020-00381 and 2021-05811), by the Carl Trygger Foundation and by the Olle Engkvist Foundation. This project has also received funding from the European Union’s Horizon Europe research and innovation program under Grant Agreement No. 101046968 (BRISQ) and support from COST Action No. CA17113 “Trapped Ions: Progress in Classical and Quantum Applications”. The authors also thank the Joint Brazilian-Swedish Research Collaboration (CAPES-STINT), grant 88887.304806/2018-00 and BR2018-8054. R.B. and C.J.V.-B thank the support from the National Council for Scientific and Technological Development (CNPq) grants 307077/2018-7, 311612/2021-0, 141247/2018-5, 409946/2018-4 and 313886/2020-2, and from the S\~ao Paulo Research Foundation (FAPESP) through Grants No.~2020/00725-9, 2019/13143-0, 2019/11999-5, 2022/00209-6, and 2018/15554-5. A.C. acknowledges support by
the S\~ao Paulo Research Foundation (FAPESP, Grants Nos. 2017/09390-7, 2022/06449-9, and 2023/07463-8) and CAPES-STINT collaboration (Grant No. 88887.649265/2021-00). A.C.S. acknowledges the support by the São Paulo Research Foundation (FAPESP) through Grants Nos. 2019/22685-1 and 2021/10224-0, and by the Proyecto Sinérgico CAM 2020 Y2020/TCS-6545 (NanoQuCo-CM) from the Comunidad de Madrid. This work has been supported by the French government, through the UCA$^\textrm{JEDI}$ Investments in the Future project managed by the National Research Agency (ANR) with the reference number ANR-15-IDEX-01. \\


\appendix

\section{Experimental setup}
\noindent A single $^{88}$Sr$^{+}$ ion is confined by a macroscopic linear Paul trap. Trapping frequencies in the radial plane - perpendicular to the trap axis - are defined by the amplitudes of the overlapping static and radio-frequency quadrupole fields; motion along the trap axis is set by the amplitude of the static axial quadrupole field and is not considered here due to the negligible coupling between axial and radial modes. Doppler cooling is performed by a single beam directed at $45^{\circ}$ to the axial mode and $60^{\circ}$ to each of the two radial modes. Carrier and sideband pulses are performed with a single beam perpendicular to the trap axis and with a $45^{\circ}$ overlap to each of the two radial modes. Fluorescence detection is performed using a photomultiplier tube (PMT). To perform coherent excitation of the radial modes, we modulate the fixed voltage applied to a pair of electrodes that run parallel to the trap axis. These are typically used to alter the horizontal position of the static quadrupole field null in order to reduce excess micromotion. 

\section{Dark and bright phononic states}
\label{appendix:bright and dark states}

\noindent We consider a two-level system encoded in the electronic states $\{\ket{\uparrow},\ket{\downarrow}\}$ of a single ion in a two-dimensional harmonic potential, with frequencies $\nu_{1}$ and $\nu_{2}$ for the modes $1$ and $2$, respectively. The two-level system has a natural transition frequency $\omega_{\mathrm{a}}$ between the ground $\ket*{\downarrow}$ and excited states $\ket*{\uparrow}$ and the quantization axis is defined by a static magnetic field applied along the trap axis ($z$ direction). The bare (non-interacting) Hamiltonian for the two modes and the ion is then given by
\begin{equation}
    \hat{H}  = \frac{1}{2}\hbar\omega_a\hat{\sigma}_z + \hbar\nu_1\left(\hat{a}^{\dagger}_1 \hat{a}_1 + \frac{1}{2}\right)+\hbar\nu_2\left(\hat{a}^{\dagger}_2 \hat{a}_2 + \frac{1}{2}\right), 
\end{equation}
where $\hat{\sigma}_{z}$ is the Pauli spin operator for the principle axis and $\hat{a}_1$($\hat{a}_{1}^{\dagger}$) and $\hat{a}_2$($\hat{a}_{2}^{\dagger}$)  are the annihilation (creation) operators for the two phonon modes.

When the k-vector of an incoming laser field has an overlap with the two harmonic oscillator modes, detuning the laser to a frequency $\omega_{m} \equiv \omega_l = \omega_{\mathrm{a}} \pm \nu_{m}$ for either mode $m\in(1,2)$ allows driving the first blue (BSB) or red (RSB) sideband transitions, such that a single phonon is added or removed when flipping the spin state. For a bichromatic laser field where each constituent signal has a well-defined frequency $\omega_m$ and phase $\phi_{m}$, assuming the conditions of the Lamb-Dicke regime are met and that the transition linewidth is small compared to the motional frequency ($\Delta\omega_a \ll \nu_{m})$, the atom-light coupling is described by  
\begin{equation}
	\hat{H}_{\mathrm{int}} = \sum_{m}
	i\hbar \Omega_{m,0} \eta_{m} \hat{\sigma}^{+} \left(\hat{a}_{m}e^{-i\nu_{m} t} + \hat{a}_{m}^{\dagger}e^{i\nu_{m} t}\right)e^{- i(\delta_{m} t +\phi_{m}) } + \mathrm{h.c.}
	, \label{Ap:Eq:Hamilt}
\end{equation}
where $\Omega_{m,0} \eta_{m}$ is the Lamb-Dicke scaled interaction strength and the detuning $\delta_{m} = \omega_{m} - \omega_{a}$. The detuning of each of the two constituent signals is controlled in order to perform BSB transitions, with $\delta_{m}=\nu_{m}$, and RSB transitions, with $\delta_{m}=-\nu_{m}$ ~\cite{Leibfried:03}. In the latter case the interaction Hamiltonian simplifies to the two-mode Jaynes-Cummings Hamiltonian (by adequately choosing the phase as $ie^{i\phi_{m} }=1$ and defining $g_m\equiv\Omega_{m,0}\eta_{m}$)
\begin{align}
	\hat{H}_{\mathrm{int}} & = \sum_{m}\hbar g_{m}\left(\hat{\sigma}^{+} \hat{a}_{m} + \hat{\sigma}^{-} \hat{a}^{\dagger}_{m}\right) \nonumber \\
	   & = \hbar g_{1}\left(\hat{\sigma}^{+} \hat{a}_{1} + \hat{\sigma}^{-} \hat{a}^{\dagger}_{1}\right)
	+
	\hbar g_{2} \left(\hat{\sigma}^{+} \hat{a}_{2} + \hat{\sigma}^{-} \hat{a}^{\dagger}_{2}\right).
\end{align}

The dark and bright phononic states for the above Hamiltonian depend on the ratio $g_{2}/g_{1}$, but in the fully symmetric case $g_{2}/g_{1}=1$ (of interest in our work) we observe that for any arbitrary motional state of this system can be written in terms of the following two-mode Fock basis
\begin{align}
    \ket*{\psi_{n}^{N}} = &\sum_{m = 0}^{N}C_{m,n}^{N}\ket*{m,N-m},\\
    C_{m, n}^N = &\sqrt{\frac{n! (N-n)!}{2^N}} \times \nonumber \\
    \quad & \sum_{q = q_\mathrm{min}}^{q_\mathrm{max}} \frac{(-1)^{m-q}\sqrt{m!(N-m)!}}{q! (n-q)! (m-q)! (N-n-m+q)!}
\end{align}

\noindent where $N$ is the total number of phonons in the system, $0 \leq n \leq N$, $q_\mathrm{min} = \min(0, n+m-N)$ and $q_\mathrm{max} = \max(n, m)$ \cite{Maximo:01}. For every value $N>1$ there exists one maximally bright and one perfectly dark basis state, corresponding to the cases when $n = N$ and $n  = 0$ respectively.

\section{Extended methods}
\subsection{Preparation of single-phonon superposition states}
\label{appendix:single phonon prep}
As mentioned in Sec.~\ref{section:quantum_section} the single-phonon superposition state is prepared with the following pulse sequence:

\vspace{-1em}

\begin{gather*}
\ket*{\downarrow,0,0}\\
\begin{tikzpicture}
    \hspace{5em}
    \draw[->] (0,0.6) -- (0,0);
    \node at (2,0.3) {$\pi/2$ pulse on BSB$_1$};
 \end{tikzpicture}\\
\frac{1}{\sqrt{2}}\{\ket*{\downarrow,0,0} + \ket*{\uparrow,1,0}\}\\
\begin{tikzpicture}
    \hspace{4.9em}
    \draw[->] (0,0.6) -- (0,0);
    \node at (2,0.3) {$\pi$ pulse on BSB$_2$};
 \end{tikzpicture}\\
\frac{1}{\sqrt{2}}\{\ket*{\uparrow,1,0} + \ket*{\uparrow,0,1}\}\\
\begin{tikzpicture}
    \hspace{5em}
    \draw[->] (0,0.6) -- (0,0);
    \node at (2,0.3) {$\pi$ pulse on carrier};
 \end{tikzpicture}\\
\frac{1}{\sqrt{2}}\{\ket*{\downarrow,1,0} + \ket*{\downarrow,0,1}\}
 \end{gather*}
 
\noindent A postselection step is performed after the second BSB pulse to mitigate any error in the state preparation due to imperfect BSB transfer efficiencies (typically $\sim95\%$). Fig.~\ref{fig:prep_single} below displays the initial Fock state distributions, extracted by simultaneously fitting RSB and BSB oscillations with the phonon occupation number $N$ as a free parameter (up to $N = 7$). 
 
\begin{figure}[h!]
    \vspace{0.5em}
    \raisebox{-2em}{\includegraphics[width=\columnwidth]{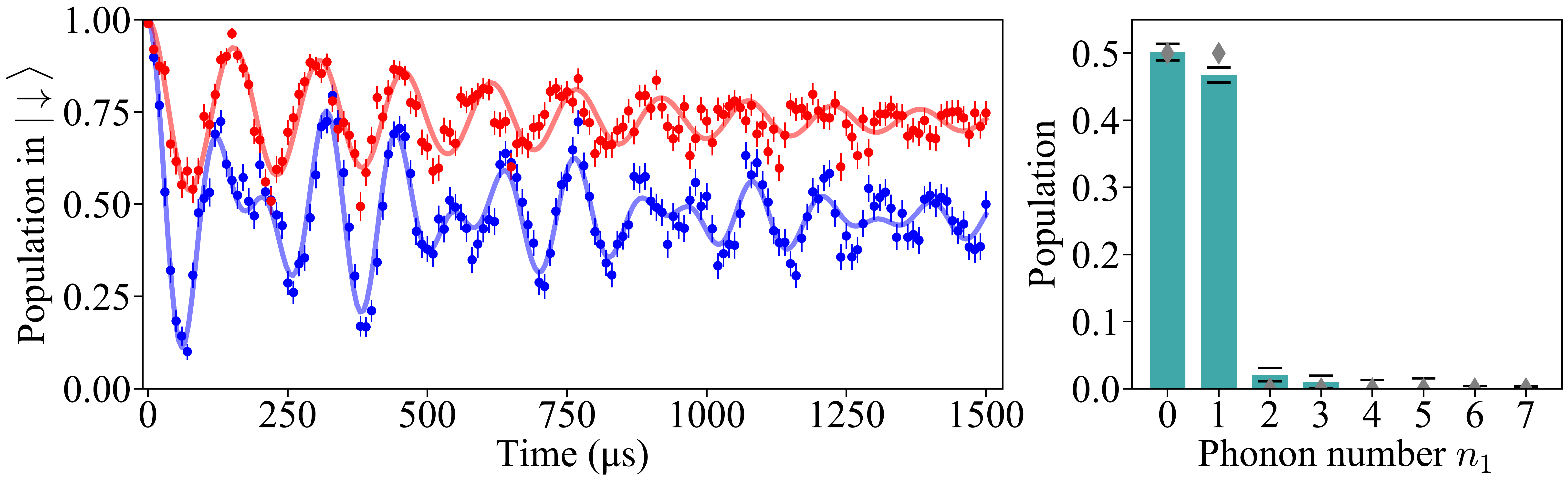}}
    \begin{picture}(0,0)
        \put(-125,75){(a)}
    \end{picture}
    \begin{picture}(0,0)
        \put(35,75){(b)}
    \end{picture}
    \includegraphics[width=\columnwidth]{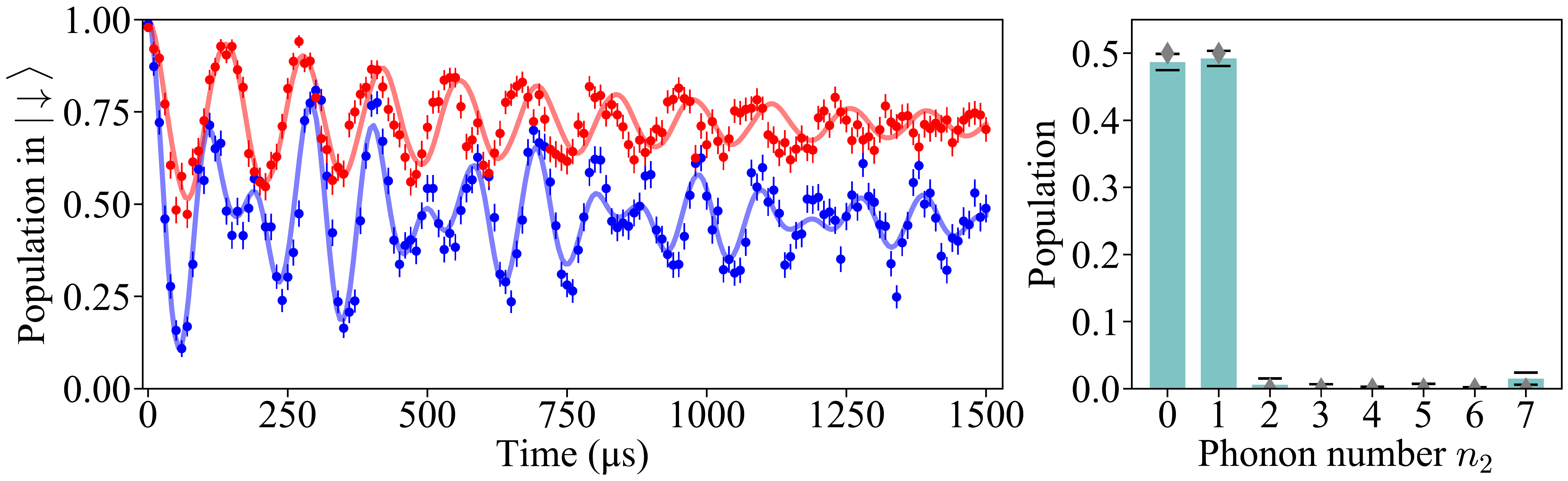}
    \begin{picture}(0,0)
        \put(-125,85){(c)}
    \end{picture}
    \begin{picture}(0,0)
        \put(35,85){(d)}
    \end{picture}
    \vspace{-1em}
    \caption{Phonon number distributions for the radial motional modes after preparation of the state $\ket*{B^1}$. (a) and (c) show the fitted RSB and BSB Rabi oscillations, error bars are from quantum projection noise ($1\sigma$ confidence intervals). (b) and (d) display the measured motional state, gray diamonds indicate the expected values assuming perfect state preparation, error bars are indicated by black lines and are obtained from fitting error.} 
    \label{fig:prep_single}
\end{figure}
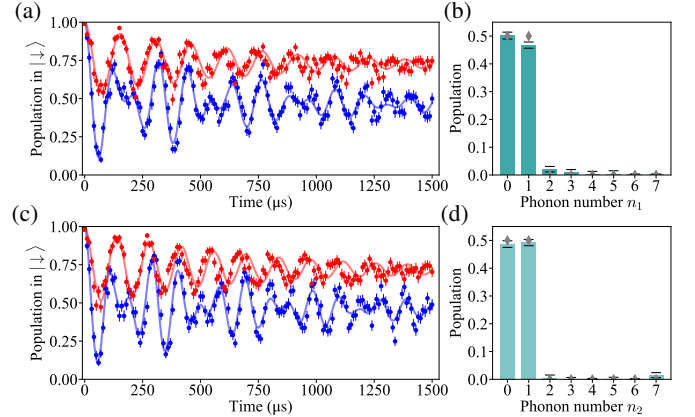

\subsection{Preparation of coherent superposition states}
\label{appendix:coherent calibration}
In order to prepare the coherent states described in Sec.~\ref{section:coherent_section} the response of the compensation electrodes to an applied modulation needs to be characterized. We first prepare a single ion in its motional ground state and subsequently monitor the magnitude of coherent displacement for each mode whilst applying a fixed modulation voltage for a variable time, the results are shown in Fig.~\ref{fig:coh_calibration}(a). The values of $\alpha$ are determined according to the method in Appendix~\ref{appendix:single phonon prep}, an example is given by the Figs.~\ref{fig:coh_calibration}(b-e) for a tickle time $t = 100\upmu$s in order to highlight that the driving field is coherently displacing the initial thermal state, rather than simply heating the ion. We also note that the initial state we refer to here is sideband cooled close to the two-mode vacuum state $\ket{0,0}$ but is more accurately represented by a very low $\Bar{n}$ thermal state due to technical imperfections of the cooling mechanism. In Fig~\ref{fig:coh_calibration}(a), a linear response in the coherent displacement is observed for both radial modes. From the fits we extract excitation rates of  $\alpha_1(t) = 6.53(11) \cdot t/\text{ms}$ for mode 1 and $\alpha_2(t) = 8.81(8) \cdot t/\text{ms}$ for mode 2.

\begin{figure}[h!]
    \includegraphics[width=0.85\columnwidth]{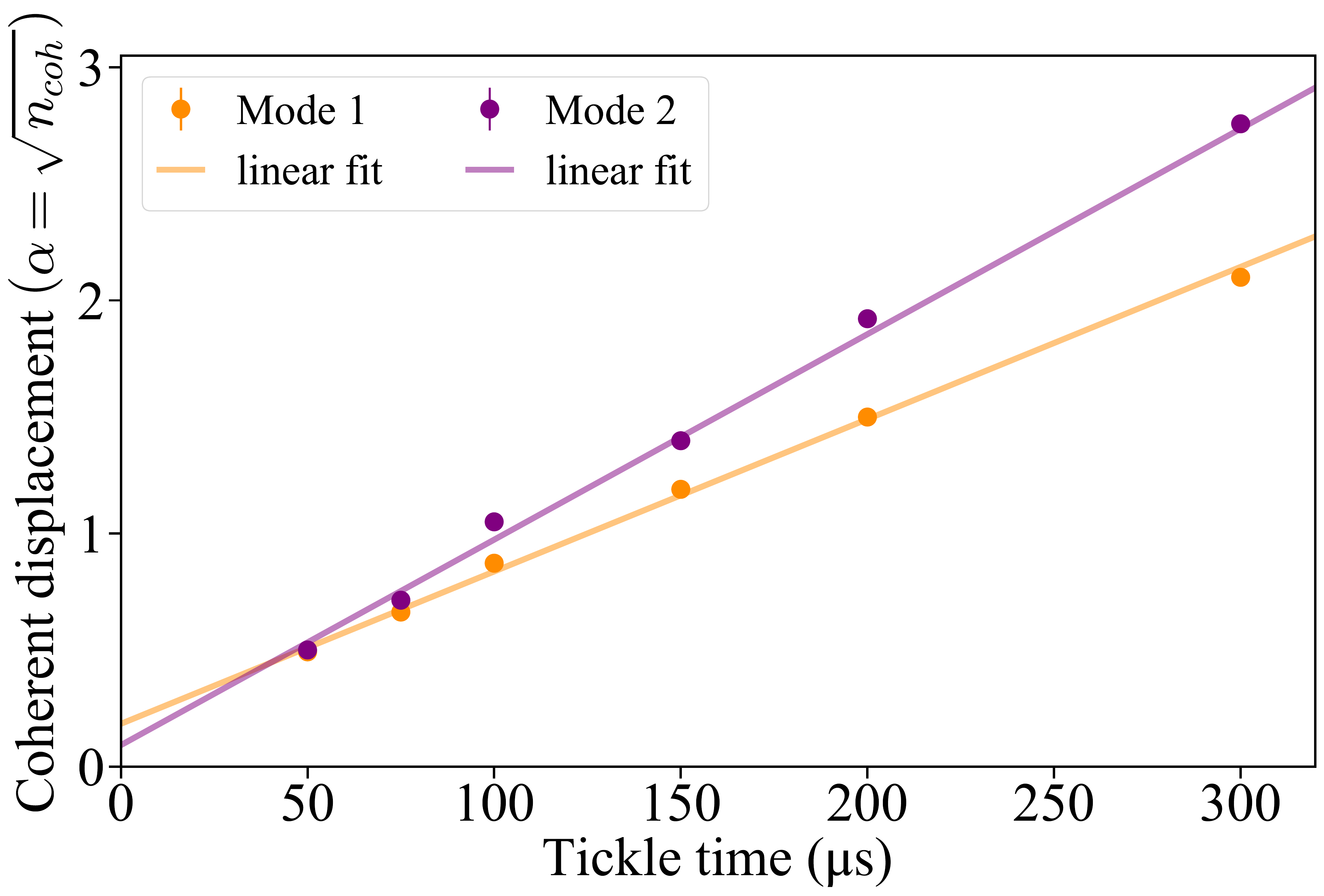}
    \begin{picture}(0,0)
        \put(-226,130){(a)}
    \end{picture}
    
    \vspace{1em}
        
    \raisebox{-2em}{\includegraphics[width=\columnwidth]{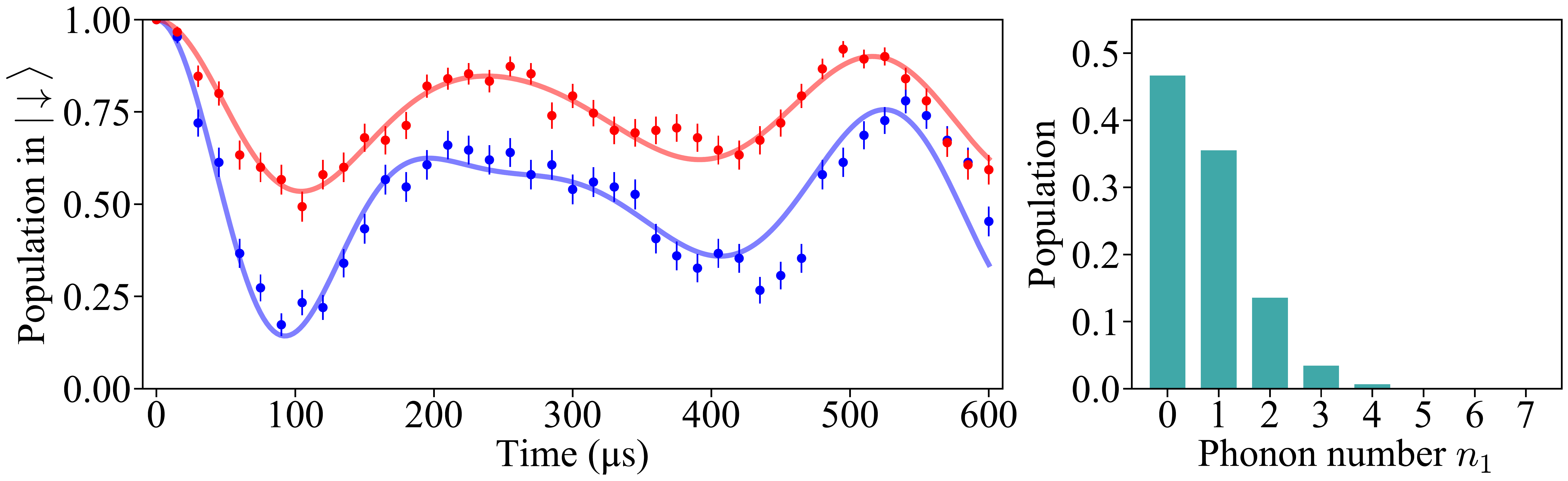}}
    \begin{picture}(0,0)
        \put(-125,75){(b)}
    \end{picture}
    \begin{picture}(0,0)
        \put(35,75){(c)}
    \end{picture}
    \includegraphics[width=\columnwidth]{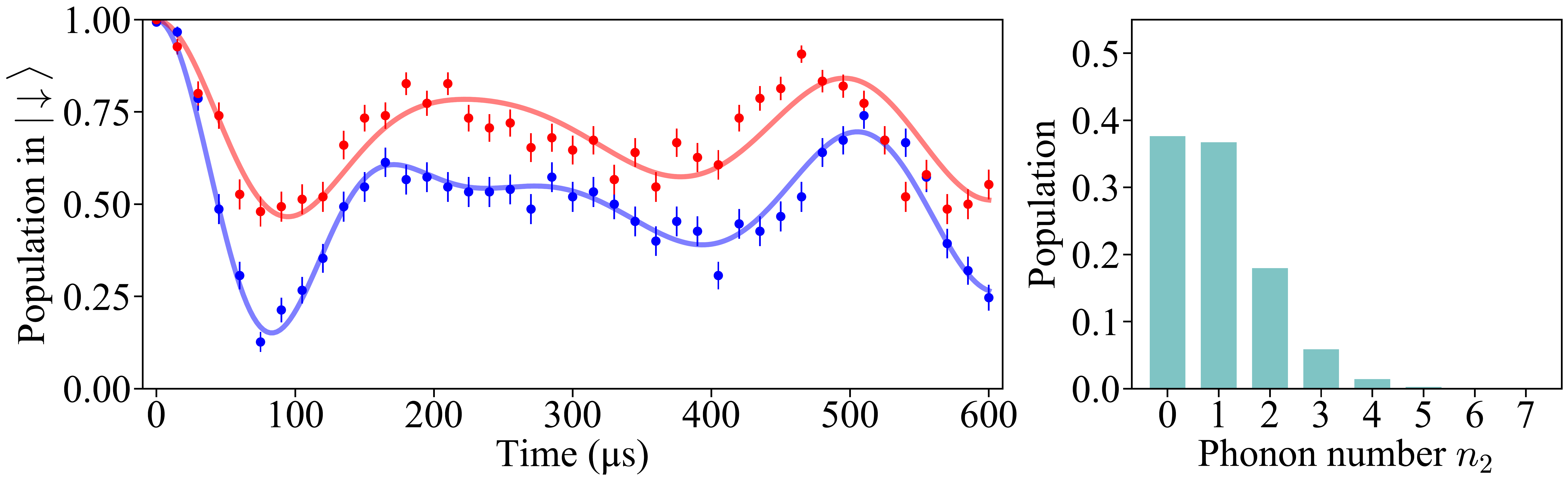}
    \begin{picture}(0,0)
        \put(-125,85){(d)}
    \end{picture}
    \begin{picture}(0,0)
        \put(35,85){(e)}
    \end{picture}
    \vspace{-1.5em}
    \caption{(a) Calibration of coherent state formation for both radial motional modes. Error bars are produced from the fitting errors of the RSB and BSB sideband Rabi oscillations used to obtain the estimates of $\alpha$; these are shown but are largely contained within the size of the data points. (b-e) Preparation of the states $\ket*{\alpha_1(t),0}$ (top) and $\ket*{0,\alpha_2(t)}$ (bottom) for a tickle time $t = 100\upmu$s. The Rabi oscillations in (b) and (d) are fit according to a coherent displacement of a well-cooled initial thermal state. Fitting error provides an error estimate for the final average phonon number $\Bar{n}$ but this is not propagated to the distributions shown in (c) and (e) to simplify the calibration process (for which we are more interested in the displacement). The phonon number distributions we obtain are indicative of coherently displaced thermal states with $\alpha \approx 1$. Error bars in (b) and (d) are from quantum projection noise ($1\sigma$ confidence intervals). }
    \label{fig:coh_calibration}
\end{figure}
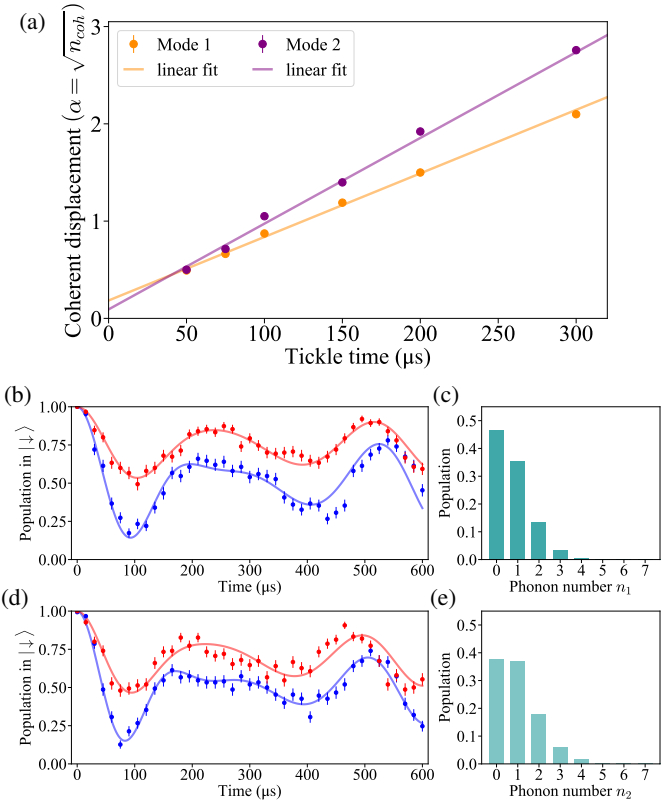

\subsection{Upsilon state formation}
\label{appendix:upsilon state formation}
To prepare the state given by Eq.~\eqref{Eq:upsilon} we initially cool the ion to its motional ground state and optically pump all population into the electronic ground state $\ket*{\downarrow} \equiv \ket*{5S_{1/2}\:, m_J=-1/2}$. During the preparation sequence we make use of carrier and sideband pulses to the excited state $\ket*{\uparrow} \equiv \ket*{4D_{5/2} \:, m_J=-5/2}$ and carrier transitions to an auxiliary state $\ket*{\uparrow'} \equiv \ket*{4D_{5/2} \:, m_J=-3/2}$. For the sequence below, sideband transitions are specified for modes 1 and 2, the carrier transition $\ket*{\downarrow} \leftrightarrow \ket*{\uparrow}$ is labelled $\textrm{CAR}_A$ and the carrier transition $\ket*{\downarrow}\leftrightarrow \ket*{\uparrow'}$ is labelled $\textrm{CAR}_B$. The full preparation of Eq.~\eqref{Eq:upsilon} is as follows: 
\begin{gather*}
\ket*{\downarrow,0,0}\\
\begin{tikzpicture}
    \hspace{5em}
    \draw[->] (0,0.8) -- (0,0);
    \node at (2,0.4) {$\pi/3$ pulse on BSB$_1$};
\end{tikzpicture}\\
\frac{\sqrt{3}}{2}\ket*{\downarrow,0,0} + \frac{1}{2}\ket*{\uparrow,1,0}\\  
\begin{tikzpicture}
    \hspace{5em}
    \draw[->] (1.65,0.8) -- (1.65,0);
    \node at (3.2,0.5) {$\pi/2.55$ pulse on};
    \node at (3.5,0.1) {BSB$_2$ with phase $\varphi_2$};
\end{tikzpicture}\\
        \frac{1}{\sqrt{2}}\ket*{\downarrow,0,0}+\frac{1}{2}\ket*{\uparrow,1,0} + e^{i\varphi_2}\frac{1}{2}\ket*{\uparrow,0,1}\\
    \begin{tikzpicture}
    \hspace{5em}
    \draw[->] (0,0.8) -- (0,0);
    \node at (2,0.4) {$\pi$ pulse on CAR$_A$};
\end{tikzpicture}\\ 
        \frac{1}{\sqrt{2}}\ket*{\uparrow,0,0}+\frac{1}{2}\ket*{\downarrow,1,0} + e^{i\varphi_2}\frac{1}{2}\ket*{\downarrow,0,1}\\
    \begin{tikzpicture}
    \hspace{5em}
    \draw[->] (0,0.8) -- (0,0);
    \node at (2,0.4) {$\pi$ pulse on CAR$_B$};
\end{tikzpicture}\\ 
        \frac{1}{\sqrt{2}}\ket*{\uparrow,0,0}+\frac{1}{2}\ket*{\uparrow',1,0} + e^{i\varphi_2}\frac{1}{2}\ket*{\uparrow',0,1}
\end{gather*}

\vspace{0.8em}

\noindent Here we perform a round of postselection to discard any results in which imperfect transfer efficiencies cause the ground state $\ket*{\downarrow}$ to remain populated. The preparation continues as:

\begin{gather*}
        \frac{1}{\sqrt{2}}\ket*{\uparrow,0,0}+\frac{1}{2}\ket*{\uparrow',1,0} + e^{i\varphi_2}\frac{1}{2}\ket*{\uparrow',0,1}\\ 
    \begin{tikzpicture}
    \hspace{5em}
    \draw[->] (0,0.8) -- (0,0);
    \node at (2,0.4) {$\pi$ pulse on CAR$_A$};
\end{tikzpicture}\\ 
        \frac{1}{\sqrt{2}}\ket*{\downarrow,0,0}+\frac{1}{2}\ket*{\uparrow',1,0} + e^{i\varphi_2}\frac{1}{2}\ket*{\uparrow',0,1}\\ 
\begin{tikzpicture}
    \hspace{5em}
    \draw[->] (1.65,0.8) -- (1.65,0);
    \node at (3.5,0.5) {$\pi/2$ pulse on BSB$_1$ };
    \node at (3.1,0.1) {with phase $\varphi_1$};
\end{tikzpicture}\\
        \frac{1}{2}\left\{\ket*{\downarrow,0,0}+e^{i\varphi_1}\ket*{\uparrow,1,0}+\ket*{\uparrow',1,0} + e^{i\varphi_2}\ket*{\uparrow',0,1} \right\}\\
\begin{tikzpicture}
    \hspace{5em}
    \draw[->] (0,0.8) -- (0,0);
    \node at (2,0.4) {$\pi$ pulse on RSB$_1$};
\end{tikzpicture}\\
        \frac{1}{2}\left\{\ket*{\downarrow,0,0}+e^{i\varphi_1}\ket*{\downarrow,1,1}+\ket*{\uparrow',1,0} + e^{i\varphi_2}\ket*{\uparrow',0,1} \right\}\\
\begin{tikzpicture}
    \hspace{5em}
    \draw[->] (0,0.8) -- (0,0);
    \node at (1.8,0.4) {$\pi/2$ pulse on CAR$_B$};
\end{tikzpicture}\\
        \begin{align*}
            \frac{1}{\sqrt{8}}\Big\{\ket*{\downarrow,0,0}&+\ket*{\uparrow',0,0}+e^{i\varphi_1}\ket*{\downarrow,1,1}+e^{i\varphi_1}\ket*{\uparrow',1,1}\\ &
            +e^{i\varphi_2}\ket*{\uparrow',0,1}+e^{i\varphi_2}\ket*{\downarrow',0,1}\\ &
            +\ket*{\uparrow',1,0}+\ket*{\downarrow',1,0}\!\Big\}
        \end{align*}
\end{gather*}

\vspace{0.5em}

\noindent Here we discard $50\%$ of the population with a second round of postselection and obtain the final state as:

\begin{gather*}
            \frac{1}{2}\left\{\ket*{\uparrow',0,0}+\ket*{\uparrow',1,0}+e^{i\varphi_2}\ket*{\uparrow',0,1} + e^{i\varphi_1}\ket*{\uparrow',1,1}\right\}\\
\begin{tikzpicture}
    \hspace{5em}
    \draw[->] (0,0.8) -- (0,0);
    \node at (2,0.4) {$\pi$ pulse on CAR$_B$};
\end{tikzpicture}\\
            \frac{1}{2}\ket*{\downarrow}\left\{\ket*{0,0}+\ket*{1,0}+e^{i\varphi_2}\ket*{0,1} + e^{i\varphi_1}\ket*{1,1}\right\}
\end{gather*}

\vspace{0.8em}

\noindent Where we can tune the phases $\varphi_1 = \varphi_2 = \varphi$ to produce the desired Upsilon state. As mentioned in the discussion of Fig.~\ref{fig:upsilon_evolution}(b), an additional phase offset $\varphi_0$ is introduced during the preparation sequence. The source of this systematic shift is not yet fully understood but is believed to be the result of static stray fields in the trap which cause a slight shift of the optical transitions. For most coherent operations the timescale is sufficiently small that a significant phase shift is not observed, however this is not the case for the Upsilon state preparation due to the lengthy pulse sequence involved.

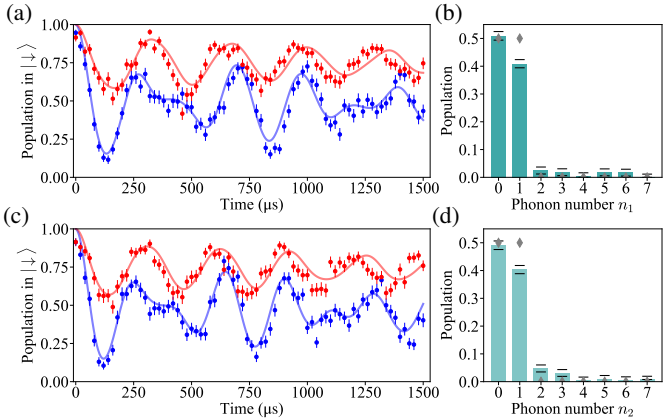
\begin{figure}[bp!]
    \vspace{0.5em}
    \raisebox{-2em}{\includegraphics[width=\columnwidth]{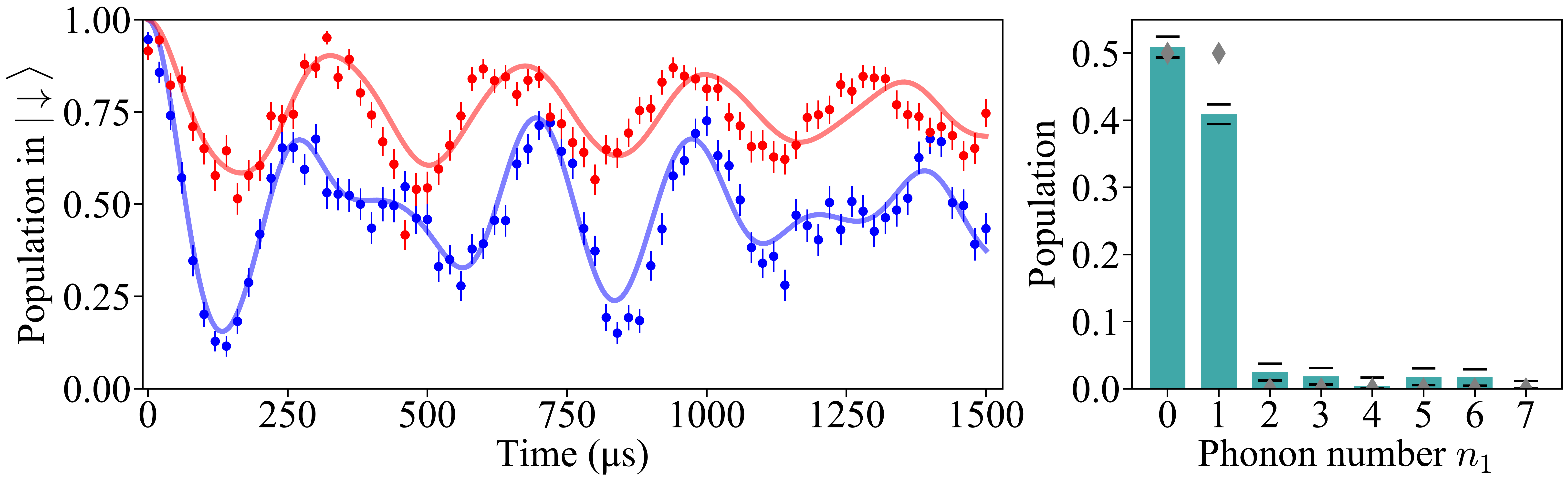}}
    \begin{picture}(0,0)
        \put(-125,75){(a)}
    \end{picture}
    \begin{picture}(0,0)
        \put(35,75){(b)}
    \end{picture}
    \includegraphics[width=\columnwidth]{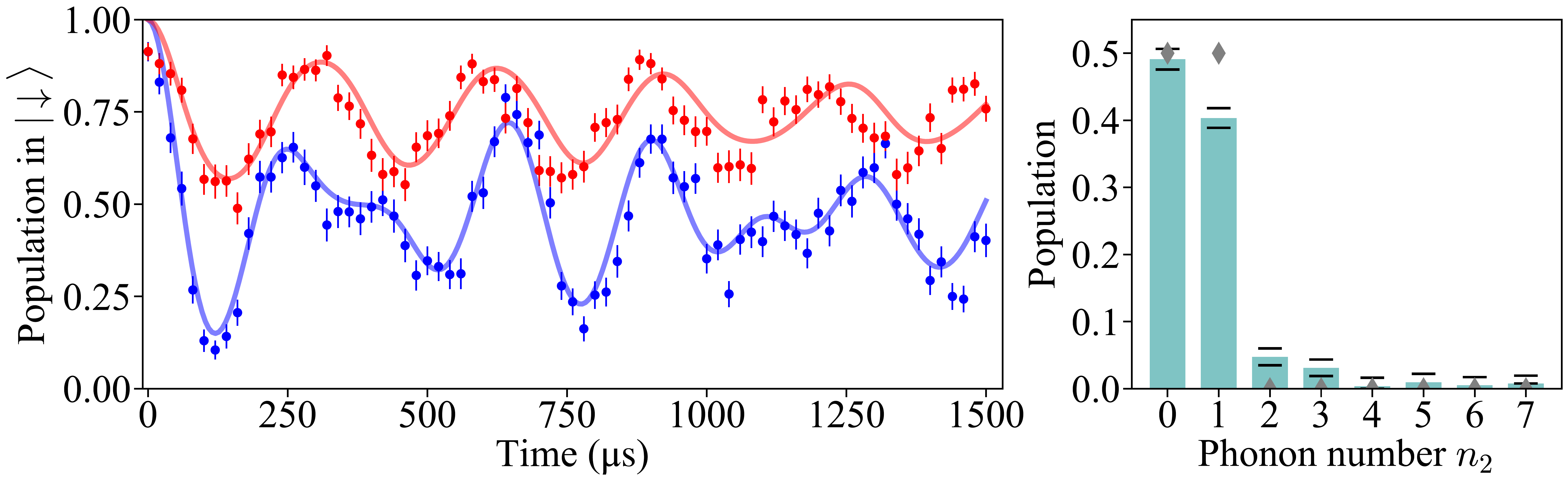}
    \begin{picture}(0,0)
        \put(-125,85){(c)}
    \end{picture}
    \begin{picture}(0,0)
        \put(35,85){(d)}
    \end{picture}
    \vspace{-1em}
    \caption{Phonon number distributions for the radial motional modes after preparation of the state $\ket*{\Upsilon_{\varphi}}$. (a) and (c) show the fitted RSB and BSB Rabi oscillations for each mode. Error bars are from quantum projection noise ($1\sigma$ confidence intervals). (b) and (d) display the measured motional state, gray diamonds indicate the expected values assuming perfect state preparation and error bars are obtained from fitting error. The overall fidelity of the state preparation is limited by technical imperfections during the extensive sequence of pulses required to obtain the state.} 
        \label{fig:prep_upsilon}
\end{figure}

The result of the state preparation is given in Fig.~\ref{fig:prep_upsilon} (obtained using the same method as in Appendix~\ref{appendix:single phonon prep}). For both modes the $n=1$ occupation probability is lower than the expected value, this infidelity is primarily attributed to the imperfect transfer efficiency of multiple pulses (particularly the RSB and BSB $\pi$ pulses). However for both modes, the discrepancy between the lowest two number states does not significantly hinder the observation of the interference effects displayed in Fig.~\ref{fig:upsilon_evolution}(a).

\section{Single phonon bright and dark states for an atom in an excited state}
\label{appendix:excited state single phonon}
\begin{figure}[h!]
        \includegraphics[width=\columnwidth]{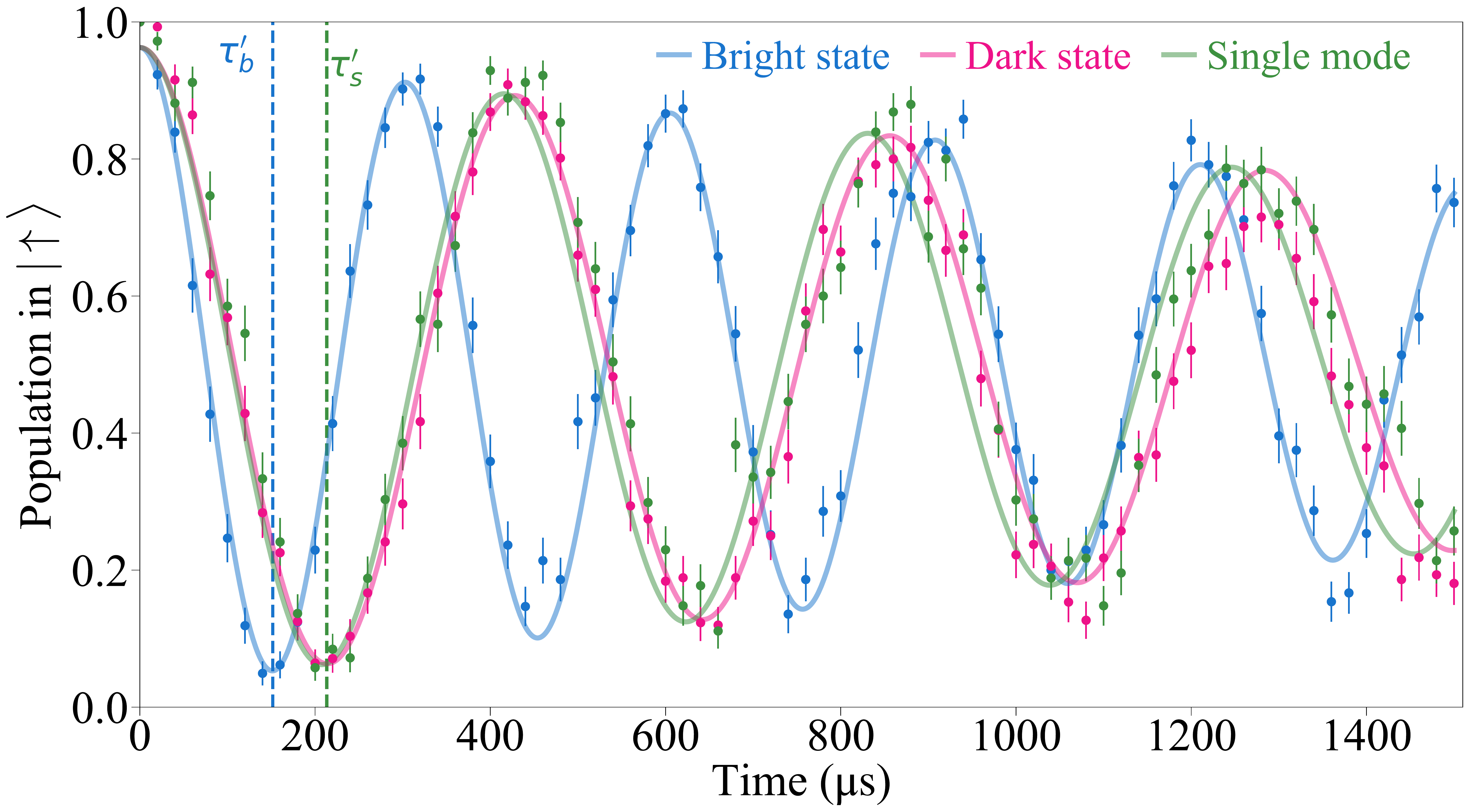}
        \caption{Rabi oscillations for bright ($\ket*{B^1}$), dark ($\ket*{D^1}$) and single-mode-excitation ($\ket*{0,1}$) states for an ion initialized in an excited electronic state ($\ket*{\uparrow}$). In this case the single-mode and dark states evolve at the same rate, which remains slower than the bright state by roughly a factor of $\sqrt{2}$. The solid lines are fits to the data produced by simulating the evolution of each state with optimized experimental parameters. Error bars indicate quantum projection noise ($1\sigma$ confidence intervals).}
        \label{fig:excitedevo}
\end{figure}

We extend our study of the single phonon bright and dark states to include interaction with an atom in its excited state. Evolution of the states from Eq.~\eqref{Eq:brightCoupling} and Eq.~\eqref{Eq:darkCoupling} as well as the single-mode-excitation state $\ket*{0,1}$ are plotted in Fig.~\ref{fig:excitedevo}. $\tau_b'$ and $\tau_s'$ (indicated by the dashed lines) are the bichromatic pulse durations which were used to obtain the results in Figs.~\ref{fig:ground_evo}(d) and \ref{fig:ground_evo}(e) respectively.   

\section{Simulations}
\label{appendix:simulations}

\begin{table*}[t!]
    \begin{tabular}{>{\rowmac}l|>{\rowmac}c|>{\rowmac}c|>{\rowmac}c|>{\rowmac}c|>{\rowmac}c<{\clearrow}c|}
    \setrow{\bfseries}Experiment & Coupling strength $g$ (kHz) & Motional dephasing $\gamma_m$ (kHz) & Electronic dephasing $\gamma_e$ (kHz) & Contrast & Offset\\ \hline     
        Fock state & 5.2 & 0 & 1.5 & 0.94 & 0.03\\
        Coherent state & 7.1 & 0.11 & 3.8 & 1 & 0\\
        Upsilon state & 7.3 & 0.28 & 2.9 & 0.68 & 0.11\\
    \end{tabular}
    \caption{Final parameters used to produce the theoretical curves in Figs~\ref{fig:ground_evo}(a), \ref{fig:coherent_evo} and \ref{fig:upsilon_evolution}(a). For the Fock and coherent states, each of the parameters was optimized by simultaneously fitting the evolution of the bright, dark and single-mode-excitation states. For the Upsilon states, the evolutions for $\varphi = 0, \varphi = \pi/2$ and $\varphi = \pi$ were fitted simultaneously.}
    \label{tab:sim_params}
\end{table*}

To simulate the dynamics including decoherence effects we solve the Lindblad master equation \cite{Petruccione:Book}
\begin{align}
    \nonumber \dot{\hat\rho}(t) =& - \frac{i}{\hbar} \left[ \hat H_{\textrm{int}}, \hat\rho(t) \right] \\+& 
    \sum_{n \in {1, 2, e}} \frac{1}{2} \left[ 2 \hat C_n \rho(t) \hat C^\dagger_n - \hat\rho(t) \hat C_n \hat C^\dagger_n - \hat C_n \hat C^\dagger_n \hat\rho(t) \right]
\end{align}
with collapse operators $\hat C_{1/2} = \sqrt{\gamma_m} \hat a_{1/2}^\dagger \hat a_{1/2}$ and $\hat C_{e} = \sqrt{\gamma_e} \hat \sigma^+ \hat \sigma\-$ describing motional and electronic coupling to the environment. The timescale for spontaneous emission is much longer ($\sim 400$ms) \cite{Letchumanan:01} than the longest operation times considered in our experiment ($\sim$ 1ms), justifying dephasing as the main decoherence mechanism in our model. The Hamiltonian $\hat{H}_{\textrm{int}}$ is then given by Eq.~\eqref{Eq:RBS}.

The time evolution was calculated using the \textit{mesolve} method of the Python package QuTip \cite{Johansson:02}. The decoherence parameters $\gamma_m, \gamma_e$, and the bare coupling strength $g$ were chosen to give the best fit with the measured data. Measurements from the same experimental run are presented in the same figure and share the same decoherence and coupling strength parameters. To account for errors in the state preparation we introduce rescaling ($A$) and offset ($C$) parameters. The optimized parameters are given in table~\ref{tab:sim_params}. 

For each model the electronic dephasing is at least an order of magnitude larger than the motional dephasing. Measurements of the coherence time of our system have shown that dephasing on the optical transition is primarily a result of laser phase fluctuations. A setup for fiber noise cancellation \cite{Ma:01} has since been implemented and should lead to a reduced rate of electronic dephasing in future experiments.

\bibliography{My_bib.bib}

\clearpage

\end{document}